\documentclass{iopart}
\usepackage{graphicx}

\usepackage{latexsym}
\usepackage{amssymb}

\jl{6}        
\eqnobysec    

\def\beq{\begin{equation}}
\def\eeq{\end{equation}}
\def\rmd{{\rm d}}

\begin{document}

\title[]
{Generalized Kerr spacetime with an arbitrary mass quadrupole moment: geometric properties vs particle motion}

\author{
Donato Bini${}^*{}^\S$, Andrea Geralico${}^\S{}^\ddag$, Orlando Luongo${}^\P{}^\S{}^\ddag$
and Hernando Quevedo${}^\flat{}^\S$
}

\address{
  ${}^*$\
Istituto per le Applicazioni del Calcolo ``M. Picone,'' CNR I-00185 Rome, Italy
}

\address{
  ${}^\S$\
  ICRA,
  University of Rome ``La Sapienza,'' I--00185 Rome, Italy
}

\address{
  $^\ddag$
  Physics Department,
  University of Rome ``La Sapienza,'' I--00185 Rome, Italy
}

\address{
$^\P$ Dipartimento di Scienze Fisiche, Universit\`a di Napoli
\lq\lq Federico II,'' Compl. Univ. di Monte S. Angelo, Edificio G, Via
Cinthia, I-80126 - Napoli, Italy 
}

\address{
  $^\flat$
Instituto de Ciencias Nucleares,
Universidad Nacional Aut\'onoma de M\'exico
}

\begin{abstract}
An exact solution of Einstein's field equations in empty space first found in 1985 by Quevedo and Mashhoon is analyzed in detail. This solution generalizes Kerr spacetime to include the case of matter with arbitrary mass quadrupole moment and is specified by three parameters, the mass $M$, the angular momentum per unit mass $a$ and the quadrupole parameter $q$.
It reduces to the Kerr spacetime in the limiting case $q=0$ and to the Erez-Rosen spacetime when the
specific angular momentum $a$ vanishes.  
The geometrical properties of such a solution are investigated.
Causality violations, directional singularities and repulsive effects occur in the region close to the source. Geodesic motion 
and accelerated motion are studied on the equatorial plane which, due to the reflection symmetry property of the solution, turns out to be also a geodesic plane. 
\end{abstract}

\pacno{04.20.Cv}

\section{Introduction}
\label{sec:int}

The problem of describing the gravitational field of astrophysical bodies is of central importance in 
general relativity, both as an issue of principle and as a foundation for explaining the results of 
observations. It is an issue of principle because general relativity is believed to be the most accurate
theory of the gravitational field. Consequently, Einstein's theory should accept the existence of exact
solutions that correctly describe the gravitational field of realistic sources. On the other hand, 
the explanation of effects observed at the astrophysical level is extremely important.

Astrophysical bodies are characterized in general by a non-spherically symmetric distribution of mass. In many cases, 
like ordinary planets and satellites,  
it is possible to neglect the deviations from spherical symmetry; it seems instead reasonable to expect that 
deviations should be taken into account in case of strong gravitational fields.

The general metric describing the gravitational field of a rotating deformed mass was found in 1991 by Quevedo and Mashhoon \cite{quev89,quevmas91} and involves an infinite set of gravitoelectric and gravitomagnetic multipoles.
This is a stationary axisymmetric solution of the vacuum Einstein's equations belonging to the class of  
Weyl-Lewis-Papapetrou \cite{weyl17,lew32,pap66} and is characterized, in general,  by the presence of a naked singularity.
To capture the main properties of the general solution \cite{quev89,quevmas91}, we concentrate for the sake of simplicity on the special case of the general solution that involves only three parameters: the mass $M$, the angular momentum per unit mass $a$ and the mass quadrupole parameter $q$ of the source.
This special case was first found by Quevedo and Mashhoon in 1985 \cite{quevmas85}.
Hereafter this solution will be denoted as the QM solution.

The corresponding line element in prolate spheroidal coordinates ($t,x,y,\phi$) with $x \geq 1$, $-1 \leq y \leq 1$ is given by \cite{ES}
\begin{eqnarray}\fl\quad
\label{metgen}
\rmd s^2&=&-f(\rmd t-\omega \rmd\phi)^2\nonumber\\
\fl\quad
&&+\frac{\sigma^2}{f}\left\{e^{2\gamma}\left(x^2-y^2\right)\left(\frac{\rmd x^2}{x^2-1}+\frac{\rmd y^2}{1-y^2}\right)+(x^2-1)(1-y^2)\rmd\phi^2\right\}\ ,
\end{eqnarray}
where $f$, $\omega$ and $\gamma$ are functions of $x$ and $y$ only and $\sigma$ is a constant.
They have the form
\begin{eqnarray}
f&=&\frac{R}{L} e^{-2qP_2Q_2}\ , \nonumber\\
\omega&=&-2a-2\sigma\frac{\mathfrak M}{R} e^{2qP_2Q_2}\ , \nonumber\\
e^{2\gamma}&=&\frac{1}{4}\left(1+\frac{M}{\sigma}\right)^2\frac{R}{x^2-1} e^{2\hat\gamma}\ ,
\end{eqnarray}
where
\begin{eqnarray}\fl\quad
\label{variousdefs}
R&=& a_+ a_- + b_+b_-\ , \qquad 
L = a_+^2 + b_+^2\ , \nonumber\\
\fl\quad
{\mathfrak M}&=&\alpha x(1-y^2)(e^{2q\delta_+}+e^{2q\delta_-}) a_+ +y(x^2-1)(1-\alpha^2e^{2q(\delta_++\delta_-)})b_+\ , \nonumber\\
\fl\quad
\hat\gamma&=&\frac12(1+q)^2 \ln\frac{x^2-1}{x^2-y^2} + 2q(1-P_2)Q_1 + q^2(1-P_2) \bigg[ (1+P_2)(Q_1^2-Q_2^2)\nonumber \\
\fl\quad
&&+\frac12(x^2-1)(2Q_2^2 - 3xQ_1Q_2 + 3 Q_0Q_2 - Q_2')\bigg] \ .
\end{eqnarray}
Here $P_l(y)$ and $Q_l(x)$ are Legendre polynomials of the first and second kind respectively. 
Furthermore
\begin{eqnarray}\fl\quad
\label{variousdefs2}
a_\pm &=& x( 1-\alpha^2e^{2q(\delta_++\delta_-)})\pm ( 1+\alpha^2e^{2q(\delta_++\delta_-)})\ , \nonumber\\
\fl\quad
b_\pm &=& \alpha y ( e^{2q\delta_+}+e^{2q\delta_-}) \mp \alpha ( e^{2q\delta_+}- e^{2q\delta_-}) \ , \nonumber\\
\fl\quad
\delta_\pm &=& \frac12\ln\frac{(x\pm y)^2}{x^2-1} +\frac32 (1-y^2\mp xy)+\frac{3}{4}[x(1-y^2) \mp y (x^2-1)]\ln\frac{x-1}{x+1}\ ,
\end{eqnarray}
the quantity $\alpha$ being a constant
\beq
\label{metquev}
\alpha=\frac{\sigma-M}{a}\ , \qquad \sigma = \sqrt{M^2-a^2}\ .
\eeq

The  Geroch-Hansen \cite{ger,hans} moments are given by
\begin{equation} 
M_{2k+1} = J_{2k}=0 \ ,  \quad k = 0,1,2,... 
\end{equation} 
\begin{equation} 
M_0 = M \ , \quad M_2 = - Ma^2 + \frac{2}{15}qM^3 \left(1-\frac{a^2}{M^2}\right)^{3/2}  \ , ... 
\label{elemm}
\end{equation} 
\begin{equation} 
J_1= Ma \ , \quad J_3 = -Ma^3 +  \frac{4}{15}qM^3 a \left(1-\frac{a^2}{M^2}\right)^{3/2}  \ , ....
\label{magmm}
\end{equation} 
The vanishing of the odd gravitoelectric ($M_n$) and even gravitomagnetic ($J_n)$ multipole moments is a consequence of the reflection symmetry of the solution about the hyperplane $y=0$, which we will refer to as \lq\lq symmetry'' (or equivalently \lq\lq equatorial'') plane hereafter. 
From the above expressions we see that $M$ is the total mass of the body, $a$ represents the 
specific angular momentum, and $q$  is related to the deviation from spherical symmetry. All higher multipole moments can be shown to depend
only on the parameters $M$, $a$, and $q$. 

We limit our analysis here to the case $\sigma>0$, i.e. $M>a$. 
In the case $\sigma=0$ the solution reduces to the extreme Kerr spacetime irrespective of the value of $q$ \cite{quevmas91}.
The case $\sigma$ complex, i.e. $a>M$, requires a different definition of the quadrupole parameter in order to have real Geroch-Hansen moments and can be better discussed by using Weyl cylindrical coordinates. We will not explore this case here.

In this work we analyze some geometric and physical properties of the QM solution.  The limiting cases contained in the general solution 
suggest that it can be used to describe the exterior asymptotically flat gravitational field of a rotating body with arbitrary quadrupole moment. This is confirmed by the analysis of the motion of  particles on the equatorial plane.
It turns out that the whole geometric structure of the QM spacetime is drastically changed in comparison with Kerr spacetime, leading to a number of previously unexplored physical effects strongly modifying the features of particle motion, especially near the gravitational source. 
In fact, the QM solution is characterized by a naked singularity at $x=1$, whose existence critically depends on the value of the quadrupole parameter $q$. In the case $q=0$ (Kerr solution) $x=1$ represents instead an event horizon. This bifurcating behaviour accounts for the above mentioned drastic changes with respect to the Kerr metric. 

Due to the very complicated form of the metric most of the analysis will be performed numerically.

\section{Limiting cases}
\label{sec:spe}

The QM solution reduces to the Kerr spacetime in the limiting case $q\rightarrow 0$ and to the Erez-Rosen spacetime when $a\rightarrow 0$.
Furthermore, it can be shown that the general form of the QM solution (see \ref{QMgeneral}) is equivalent, up to a coordinate transformation, to the exterior vacuum Hartle-Thorne solution once linearized to first order in the quadrupole parameter and to second order in the rotation parameter.

\subsection{Kerr solution}

For vanishing quadrupole parameter we recover the Kerr solution, with functions
\begin{eqnarray}\fl\quad
f_K&=&\frac{c^2x^2+d^2y^2-1}{(cx+1)^2+d^2y^2}\ , \quad 
\omega_K=2a\frac{(cx+1)(1-y^2)}{c^2x^2+d^2y^2-1}\ , \nonumber\\
\fl\quad
\gamma_K&=&\frac12\ln\left(\frac{c^2x^2+d^2y^2-1}{c^2(x^2-y^2)}\right)\ , 
\end{eqnarray}
where 
\beq
c=\frac{\sigma}{M}\ , \quad d=\frac{a}{M}\ , \quad c^2+d^2=1\ ,
\eeq
so that $\alpha=(c-1)/d$.
Transition of this form of Kerr metric to the more familiar one associated with Boyer-Lindquist coordinates is accomplished by the map 
\beq
\label{trasftoBL}
x=\frac{r-M}{\sigma}\ , \qquad 
y=\cos\theta\ ,
\eeq
so that $x=1$ corresponds to the outer horizon $r=r_+=M+\sigma$.

To first order in $q$ the QM solution becomes 
\begin{eqnarray}\fl\quad
f&=&f_K-2q[P_2Q_2f_K+\Lambda_+\delta_++\Lambda_-\delta_-]+O(q^2)\ , \nonumber\\
\fl\quad
\omega&=&\omega_K+2acq\left[2P_2Q_2\frac{c(x^2-y^2)+x(1-y^2)}{c^2x^2+d^2y^2-1}+\Psi_+\delta_++\Psi_-\delta_-\right]+O(q^2)\ , \nonumber\\
\fl\quad
\gamma&=&\gamma_K+q\left[2(1-P_2)Q_1+\ln\left(\frac{x^2-1}{x^2-y^2}\right)\right.\nonumber\\
\fl\quad
&&\left.+\alpha d\frac{c(x^2-y^2)+1-y^2}{c^2x^2+d^2y^2-1}(\delta_++\delta_-)\right]+O(q^2)\ , 
\end{eqnarray}
where 
\begin{eqnarray}\fl\quad
\Lambda_\pm&=&d^2(x\pm y)\frac{[c(x\mp y)+1]^2-y^2}{[(cx+1)^2+d^2y^2]^2}\ , \nonumber\\
\fl\quad
\Psi_\pm&=&\frac{x\pm y}{(c^2x^2+d^2y^2-1)^2}\{cd^2(x\mp y)^2(1\pm xy)+c(x^2-1)(1\mp xy)\nonumber\\
\fl\quad
&&+(x\mp y)[c^2(x^2-1)+d^2(1-y^2)]\}\ .
\end{eqnarray}
This approximate metric could be used to describe the exterior field of an arbitrarily rotating mass source with a small quadrupole moment. 
The lowest Geroch-Hansen multipole moments in this case coincide with those of the exact solution as given in Eqs. (\ref{elemm}) and (\ref{magmm}). Differences will appear in higher moments where all terms containing $q^2$ and higher exponents must be neglected. 

It is interesting to mention that in the limiting case $a\rightarrow M$, the QM metric leads to the extreme Kerr black hole solution, regardless
of the value of the quadrupole $q$ \cite{quevmas85}. This can be easily seen both at the level of the multipole moments (\ref{elemm}) and (\ref{magmm}) and by directly computing the limiting metric.
The latter determines in turn the limit of applicability of the QM solution, since for values in the range $a/M>1$ the multipole moments and the metric both become complex.
As stated in the Introduction we will not consider such a situation here.

\subsection{Erez-Rosen solution}

Similarly, for vanishing rotation parameter we recover the Erez-Rosen solution \cite{erez,novikov,young}. 
It is a solution of the static Weyl class of solutions (i.e. $\omega=0$) with functions 
\beq
f_{ER}=\frac{x-1}{x+1}e^{-2qP_2Q_2}\ , \qquad 
\gamma_{ER}=\hat\gamma\ ,
\eeq
which reduce to 
\beq
\label{schw}
f_S=\frac{x-1}{x+1}\ , \qquad
\gamma_S=\frac12\ln\left(\frac{x^2-1}{x^2-y^2}\right)
\eeq
when $q=0$, corresponding to the Schwarzschild solution. 

To first nonvanishing order in $a/M$ we find
\begin{eqnarray}\fl\quad
\label{asmall}
f&=&f_{ER}\left\{
1+\frac{1}{2(x^2-1)(x+1)}\left[
(x+y)(y-1)e^{4q\delta_+}-(x-y)(y+1)e^{4q\delta_-}\right.\right.\nonumber\\
\fl\quad
&&\left.\left.-2(x^2-y^2)e^{2q(\delta_++\delta_-)}
\right]\left(\frac{a}{M}\right)^2
\right\}
+O[(a/M)^4]\ , \nonumber\\
\fl\quad
\omega&=&-2M\left\{
1+\frac{e^{2qP_2Q_2}}{2(x-1)}\left[
(x+y)(y-1)e^{2q\delta_+}-(x-y)(y+1)e^{2q\delta_-}\right]
\right\}\frac{a}{M}\nonumber\\
\fl\quad
&&+O[(a/M)^3]\ , \nonumber\\
\fl\quad
\gamma&=&\gamma_{ER}+\frac14\left\{
1-\frac{1}{2(x^2-1)}\left[
(1-y^2)(e^{4q\delta_+}+e^{4q\delta_-})\right.\right.\nonumber\\
\fl\quad
&&\left.\left.+2(x^2-y^2)e^{2q(\delta_++\delta_-)}\right]
\right\}\left(\frac{a}{M}\right)^2+O[(a/M)^4]\ .
\end{eqnarray}
This approximate solution can be interpreted as a generalization of the Lense-Thirring spacetime \cite{lt18,hmt}, which is obtained in the limit $q\rightarrow 0$ by retaining terms up to the linear order in $a/M$. 
Consequently, the approximate solution (\ref{asmall}) could be used to investigate the exterior gravitational field of slowly rotating deformed bodies. A similar approximate solution, accurate to second order in the rotation parameter and to first order in the quadrupole moment, was found long ago by Hartle and Thorne \cite{ht67}.

\subsection{Hartle-Thorne solution}

The Hartle-Thorne metric describing the exterior field of a slowly rotating slightly deformed object is given by 
\begin{eqnarray}\fl
\label{HTmet}
\rmd s^2&=&-\left(1-\frac{2{\mathcal M}}{R}\right)\left[
1+2k_1P_2(\cos\Theta)+2\left(1-\frac{2{\mathcal M}}{R}\right)^{-1}\frac{{\mathcal J}^2}{R^4}(2\cos^2\Theta-1)\right]\rmd t^2\nonumber\\
\fl
&&+\left(1-\frac{2{\mathcal M}}{R}\right)^{-1}\left[
1-2k_2P_2(\cos\Theta)-2\left(1-\frac{2{\mathcal M}}{R}\right)^{-1}\frac{{\mathcal J}^2}{R^4}\right]\rmd R^2\nonumber\\
\fl
&&+R^2(\rmd\Theta^2+\sin^2\Theta\rmd\phi^2)[1-2k_3P_2(\cos\Theta)]-4\frac{{\mathcal J}}{R}\sin^2\Theta\rmd t\rmd\phi\ ,
\end{eqnarray}
where 
\begin{eqnarray}
k_1&=&\frac{{\mathcal J}^2}{{\mathcal M}R^3}\left(1+\frac{{\mathcal M}}{R}\right)-\frac58\frac{{\mathcal Q}-{\mathcal J}^2/{\mathcal M}}{{\mathcal M}^3}Q_2^2\left(\frac{R}{{\mathcal M}}-1\right)\ , \nonumber\\
k_2&=&k_1-\frac{6{\mathcal J}^2}{R^4}\ , \nonumber\\
k_3&=&k_1+\frac{{\mathcal J}^2}{R^4}-\frac54\frac{{\mathcal Q}-{\mathcal J}^2/{\mathcal M}}{{\mathcal M}^2R}\left(1-\frac{2{\mathcal M}}{R}\right)^{-1/2}Q_2^1\left(\frac{R}{\mathcal M}-1\right)\ . 
\end{eqnarray}
Here $Q_l^m$ are the associated Legendre functions of the second kind and the constants ${\mathcal M}$, ${\mathcal J}$ and ${\mathcal Q}$ are the total mass, angular momentum and mass quadrupole moment of the rotating star respectively.

It is interesting to find out the connection between the Hartle-Thorne solution and the QM solution in the appropriate limit.
To this end it is necessary to start with the general form of the QM solution as given in \ref{QMgeneral} containing an additional parameter, the Zipoy-Voorhees \cite{zipoy,voorhees} constant parameter $\delta$.
For our purposes it is convenient to set such a parameter as $\delta=1+sq$, where $s$ is a real number.
For $s=0$, i.e. $\delta=1$, we recover the solution (\ref{metgen})--(\ref{metquev}).
To first order in the quadrupole parameter $q$ and to second order in the rotation parameter $a/M$ the metric functions turn out to be 
\begin{eqnarray}\fl\qquad
f&\simeq&\frac{x-1}{x+1}\left[1-q\left(2P_2Q_2-s\ln\frac{x-1}{x+1}\right)\right]-\frac{x^2+x-2y^2}{(x+1)^3}\left(\frac{a}{M}\right)^2\ , \nonumber\\
\fl\qquad
\omega&\simeq&2M\frac{1-y^2}{x-1}\left(\frac{a}{M}\right)\ , \nonumber\\
\fl\qquad
\gamma&\simeq&\hat\gamma(1+2sq)-\frac12\frac{1-y^2}{x^2-1}\left(\frac{a}{M}\right)^2\ ,
\end{eqnarray}
where $\hat\gamma$ is defined in Eq. (\ref{variousdefs}) and terms of the order of $q(a/M)$ have also been neglected.

Introduce first Boyer-Lindquist coordinates $(t,r,\theta,\phi)$ through the transformation 
\beq
t=t\ , \qquad
x=\frac{r-M}{\sigma}\ , \qquad 
y=\cos\theta\ , \qquad
\phi=\phi\ . 
\eeq
Then set $s=-1$ and $M=M'(1+q)$.
The further transformation 
\beq
t=t\ , \qquad
r=r(R,\Theta)\ , \qquad 
\theta=\theta(R,\Theta)\ , \qquad
\phi=\phi\ 
\eeq
with 
\begin{eqnarray}\fl
r&=&R+M'q+\frac32M'q\sin^2\Theta\left[\frac{R}{M'}-1+\frac12\frac{R^2}{M'^2}\left(1-\frac{2M'}{R}\right)\ln\left(1-\frac{2M'}{R}\right)\right]\nonumber\\
\fl
&&-\frac{a^2}{2R}\left[\left(1+\frac{2M'}{R}\right)\left(1-\frac{M'}{R}\right)-\cos^2\Theta\left(1-\frac{2M'}{R}\right)\left(1+\frac{3M'}{R}\right)\right]\ , \nonumber\\
\fl
\theta&=&\Theta-\sin\Theta\cos\Theta\left\{\frac32q\left[2+\left(\frac{R}{M'}-1\right)\ln\left(1-\frac{2M'}{R}\right)\right]+\frac{a^2}{2R^2}\left(1+\frac{2M'}{R}\right)\right\}
\end{eqnarray}
finally gives the mapping between the general form of QM solution and the Hartle-Thorne metric (\ref{HTmet}) with parameters 
\beq
{\mathcal M}=M'=M(1-q)\ , \quad 
{\mathcal J}=-Ma\ , \quad 
{\mathcal Q}=\frac{{\mathcal J}^2}{M}+\frac45M^3q\ .
\eeq
Note that the previous transformation is obtained simply by combining the corresponding transformation from Kerr to Hartle-Thorne solution as given by Hartle and Thorne themselves in \cite{ht67} and that from Erez-Rosen to Hartle-Thorne solution as found by Mashhoon and Theiss in \cite{mt91}.

\section{Geometric properties of the solution}
\label{sec:geom}

\begin{figure} 
\typeout{*** EPS figure ergo_k}
\begin{center}
\includegraphics[scale=0.4]{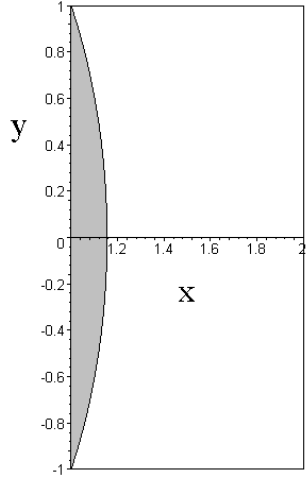}
\end{center}
\caption{The shape of the ergoregion is shown in the case of vanishing quadrupole parameter $q=0$ (Kerr spacetime) for the choice $a/M=0.5$ in prolate spheroidal coordinates. 
}
\label{fig:ergo_k}
\end{figure}

\begin{figure} 
\typeout{*** EPS figure ergo1to4}
\begin{center}
$\begin{array}{cc}
\includegraphics[scale=0.4]{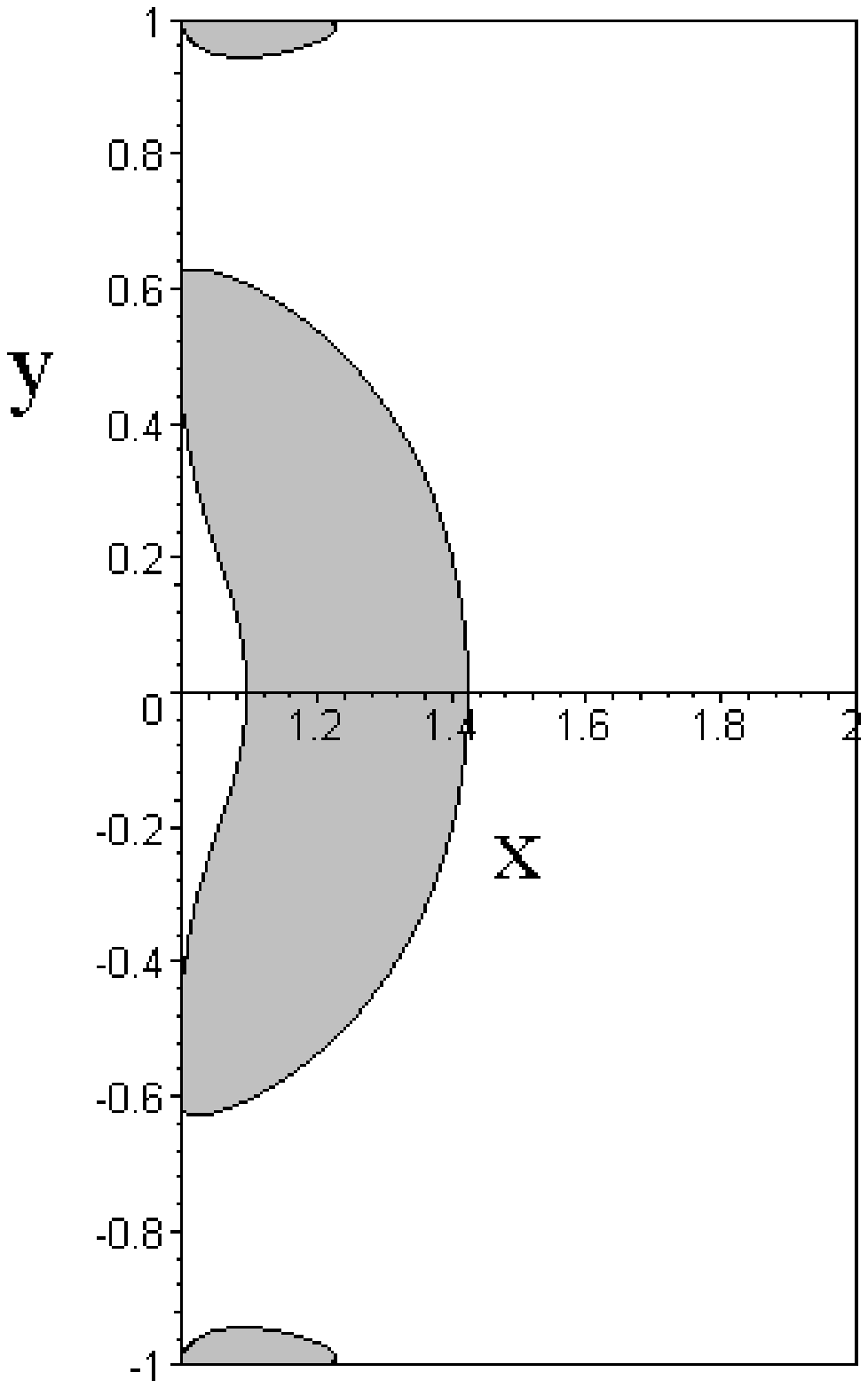}&\quad
\includegraphics[scale=0.4]{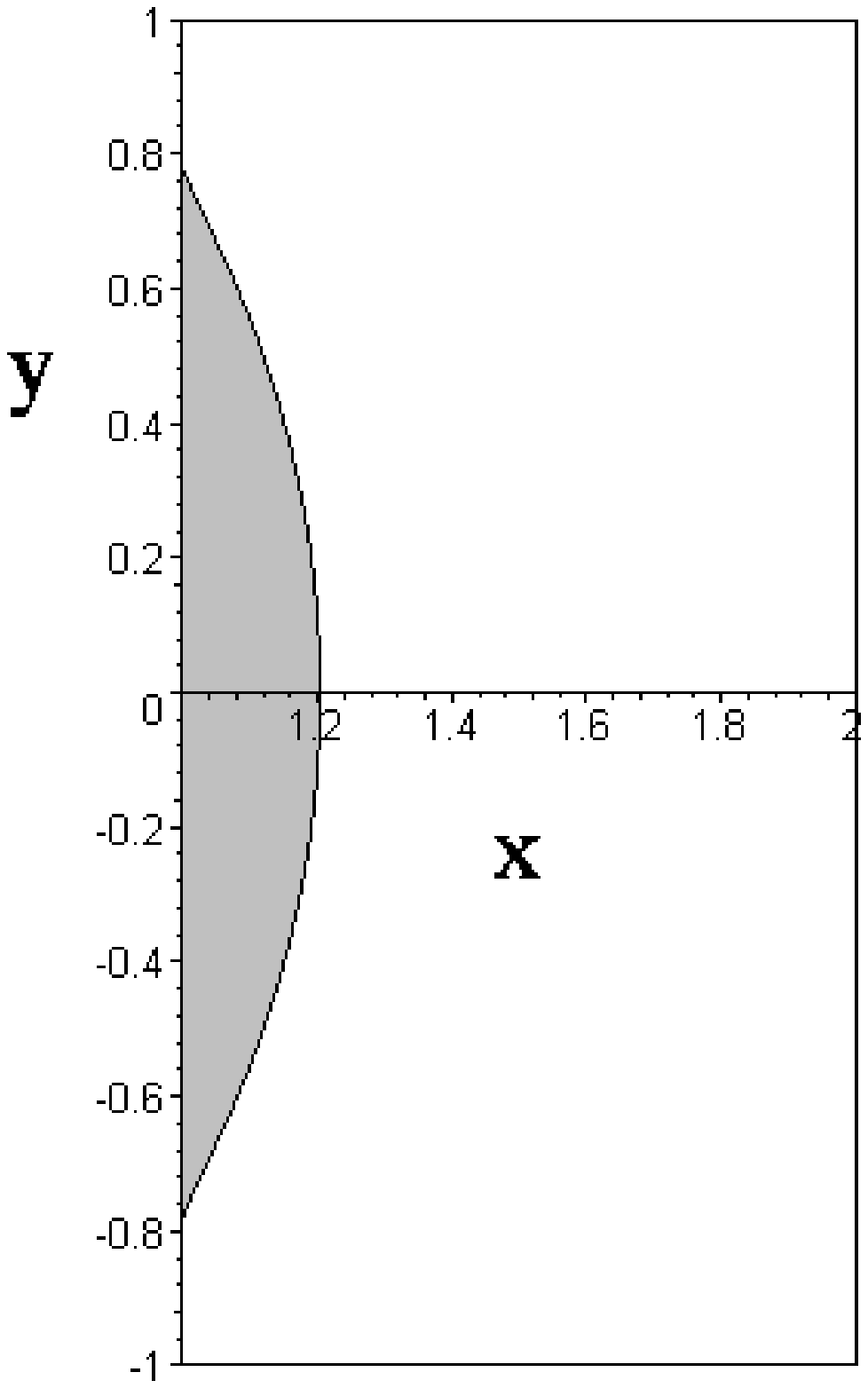}\\[.4cm]
\quad\mbox{(a)}\quad &\quad \mbox{(b)}\\[.6cm]
\includegraphics[scale=0.4]{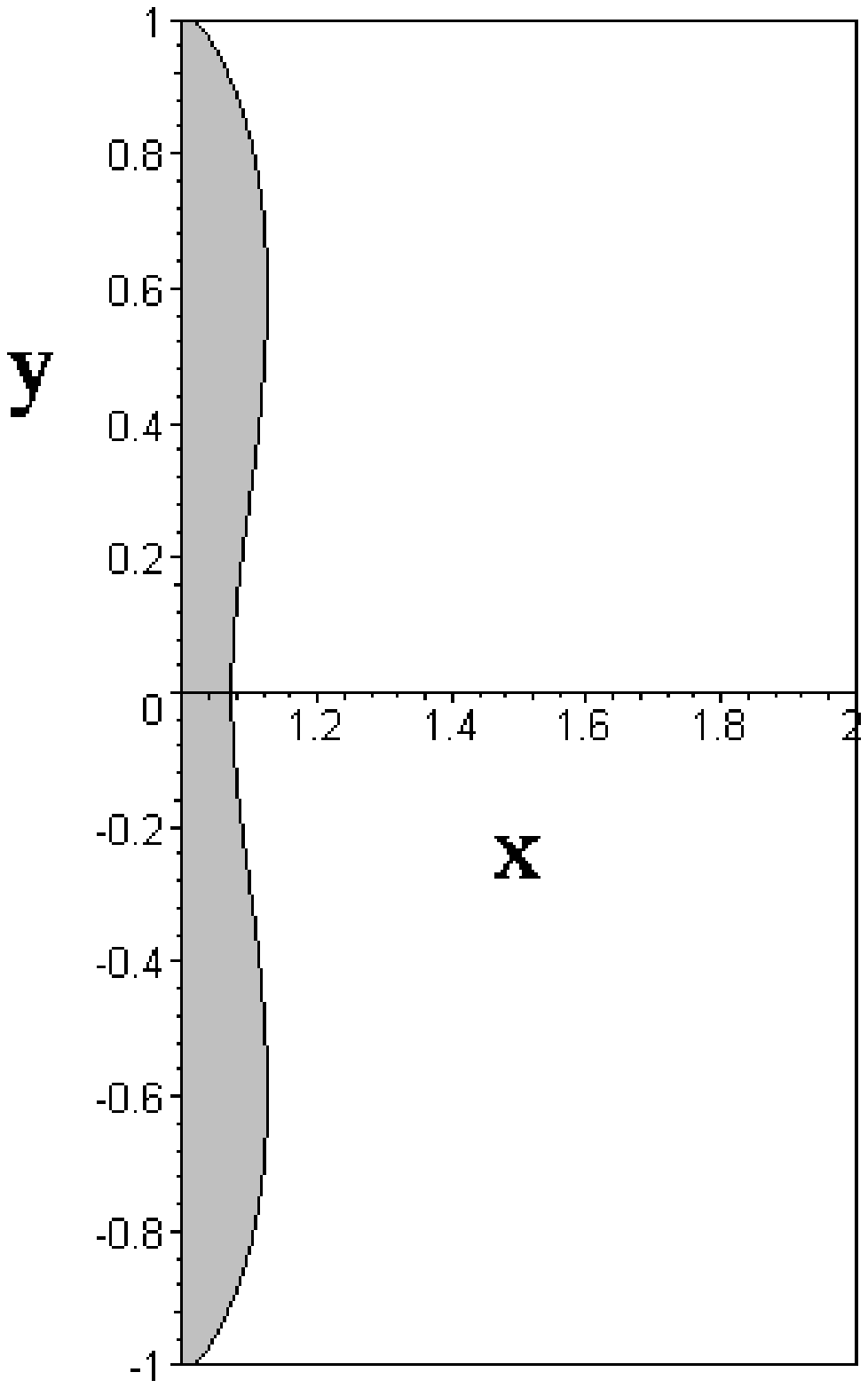}&\quad
\includegraphics[scale=0.4]{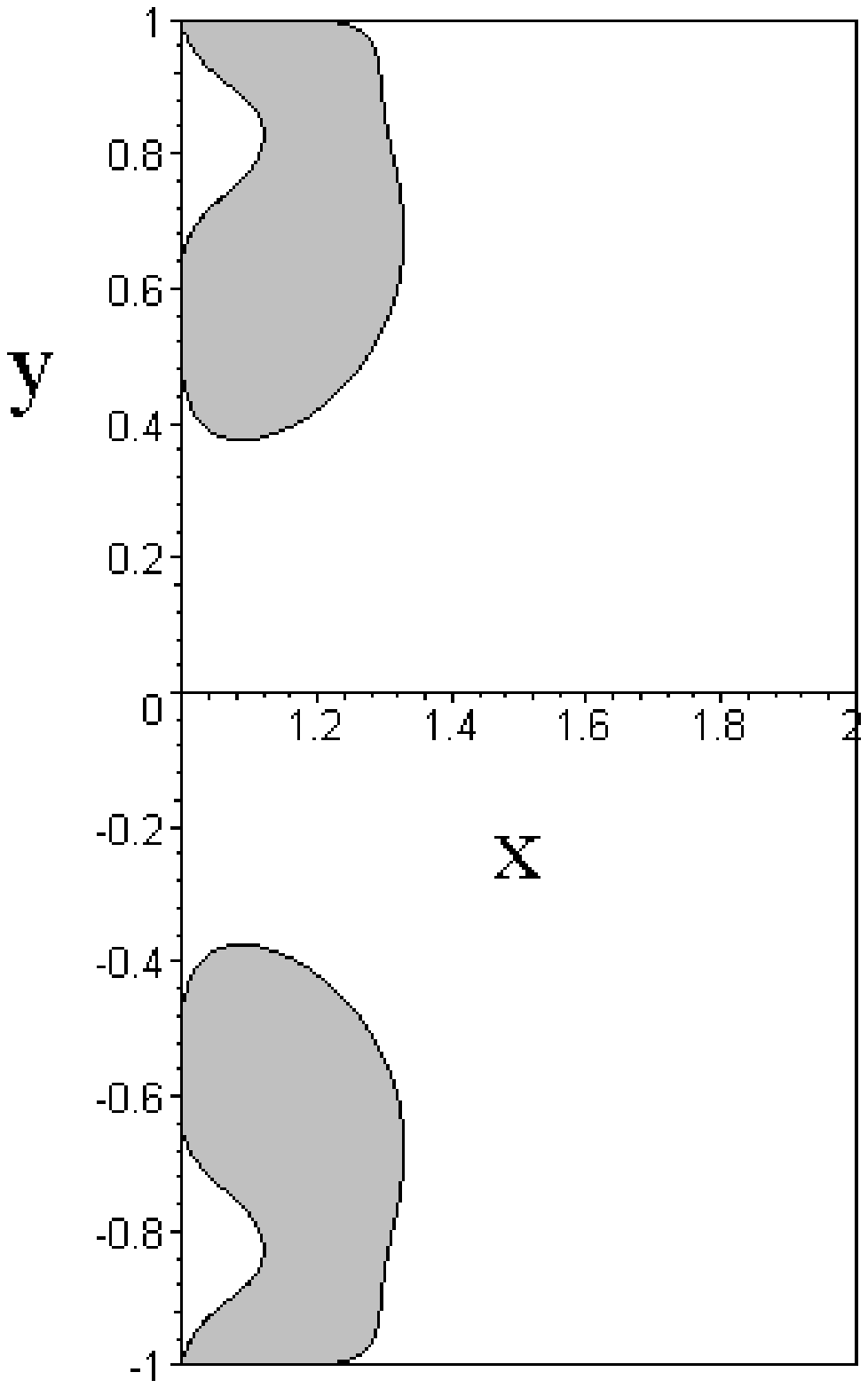}\\[.4cm]
\quad\mbox{(c)}\quad &\quad \mbox{(d)}
\end{array}$\\
\end{center}
\caption{The shape of the ergoregion is shown for $a/M=0.5$ and different values of the quadrupole parameter: $q=[-10,-1,1,10]$, from (a) to (d) respectively. 
For $q=-1$ and $q=1$ the shape is similar to the Kerr case, i.e. the regions are simply-connected. 
}
\label{fig:ergo1to4}
\end{figure}

The solution admits two Killing vectors associated with translation in time and rotation about the symmetry axis.
The timelike Killing vector $\partial_t$ changes its causality property when $f=0$, defining a hypersurface which in the Kerr limiting case is called ergosurface as the boundary of the ergosphere (or equivalently ergoregion).
Using this terminology also in this case we compare in Figs.~\ref{fig:ergo_k} and~\ref{fig:ergo1to4} the ergoregions of Kerr solution and QM solution (for different values of the quadrupole parameter) respectively.
The situation is completely different.
In fact, while in the Kerr case such a hypersurface is the boundary of a simply-connected domain, in the case of the QM solution this property is no more true, as soon as the magnitude of the quadrupole parameter exceeds a certain critical value (for example, for the choice $a/M=0.5$ the range of Kerr-like behaviour corresponds to $-1\lesssim q\lesssim 1.21$).

\begin{figure} 
\typeout{*** EPS figure ctc12}
\begin{center}
$\begin{array}{cc}
\includegraphics[scale=0.4]{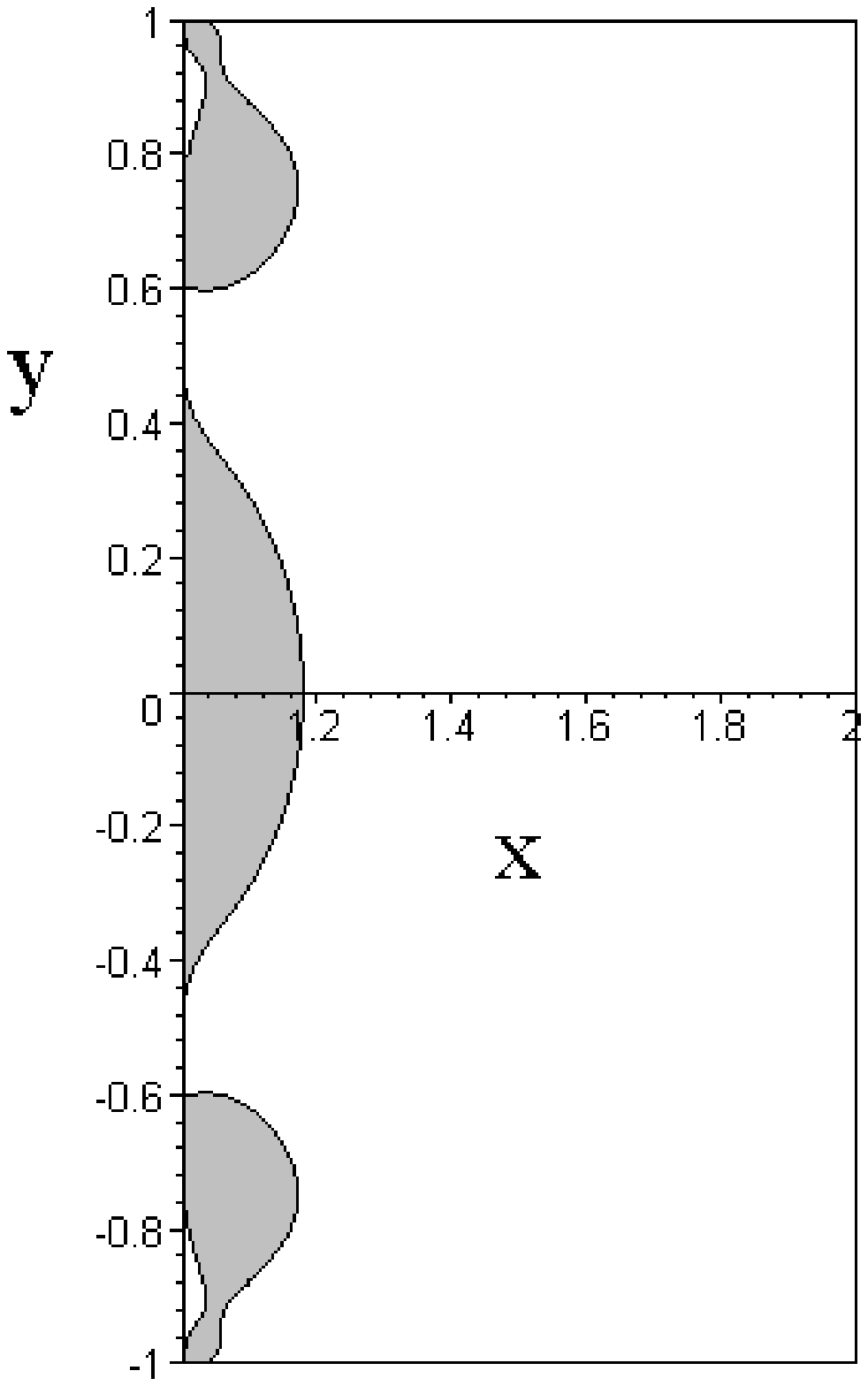}&\quad
\includegraphics[scale=0.4]{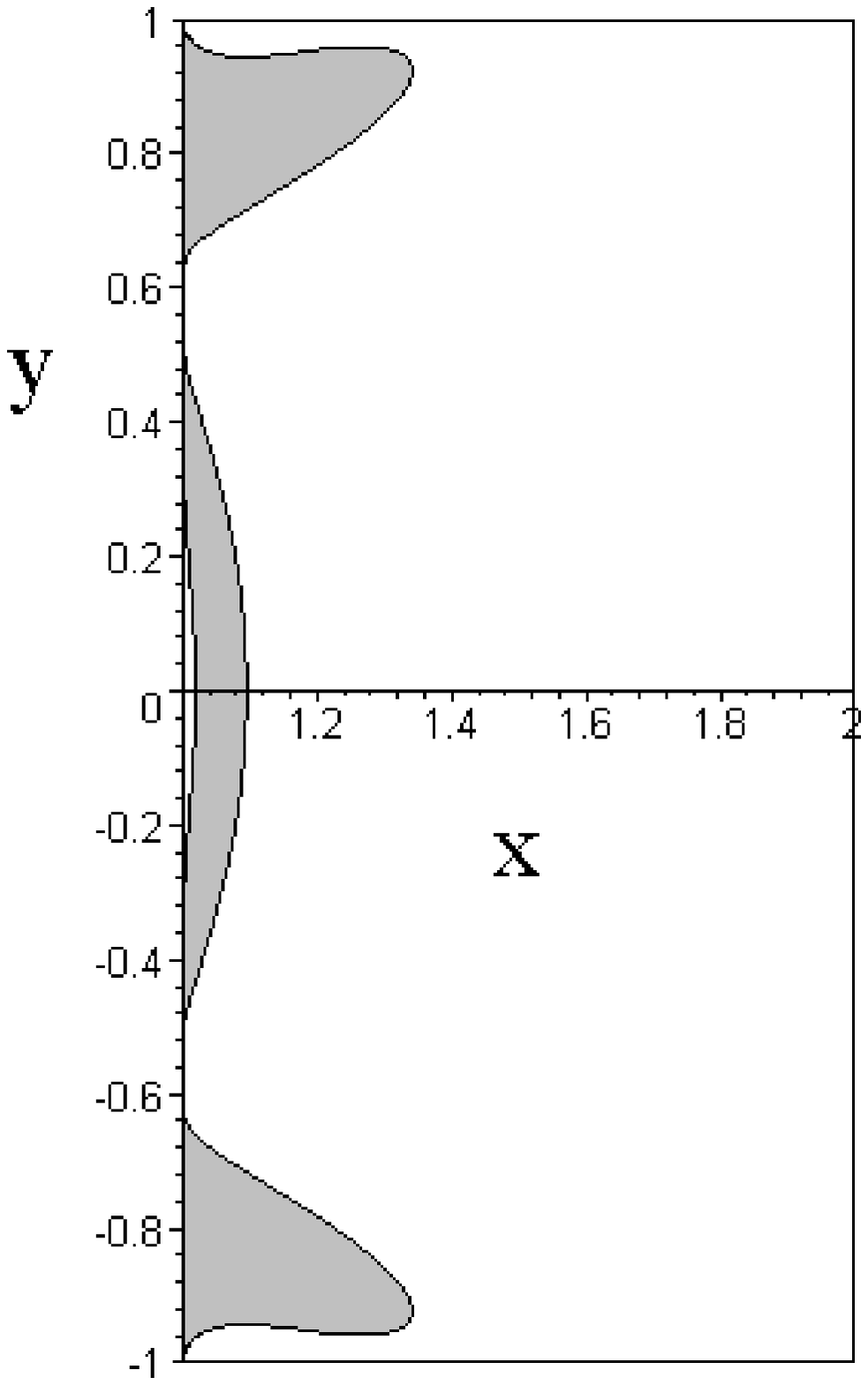}\\[.4cm]
\quad\mbox{(a)}\quad &\quad \mbox{(b)}
\end{array}$\\
\end{center}
\caption{The regions where the metric component $g_{\phi\phi}$ changes its sign are shown for $a/M=0.5$ and different values of the quadrupole parameter: (a) $q=10$ and (b) $q=-10$.
The existence of closed timelike curves is allowed there.
}
\label{fig:ctc12}
\end{figure}

The spacelike Killing vector $\partial_\phi$ also changes its causality condition in some regions, leading to the existence of closed timelike curves in the QM solution.
These regions are depicted in Fig.~\ref{fig:ctc12} for different values of the quadrupole parameter.

\begin{figure} 
\typeout{*** EPS figure invar}
\begin{center}
$\begin{array}{cc}
\includegraphics[scale=0.35]{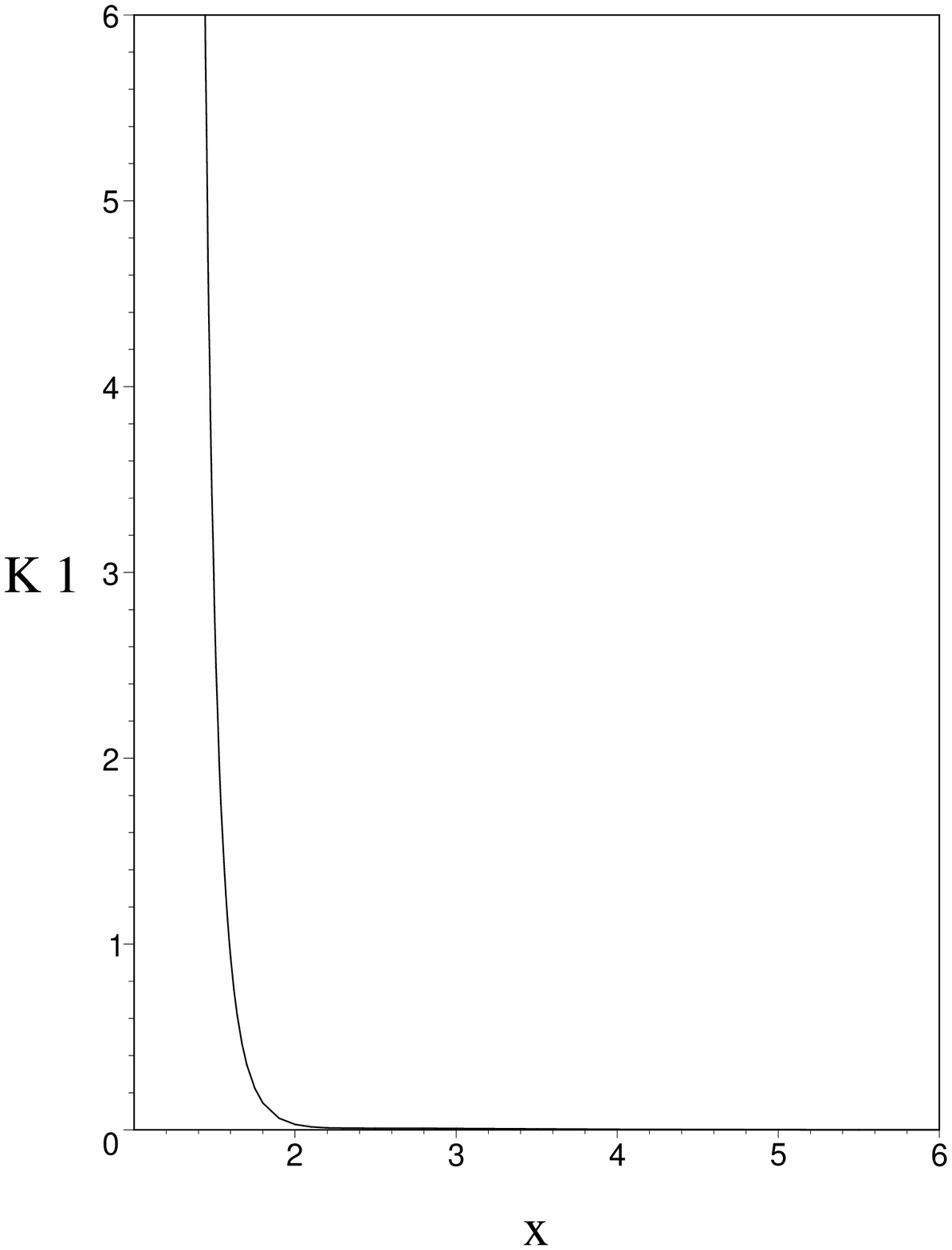}&\quad
\includegraphics[scale=0.35]{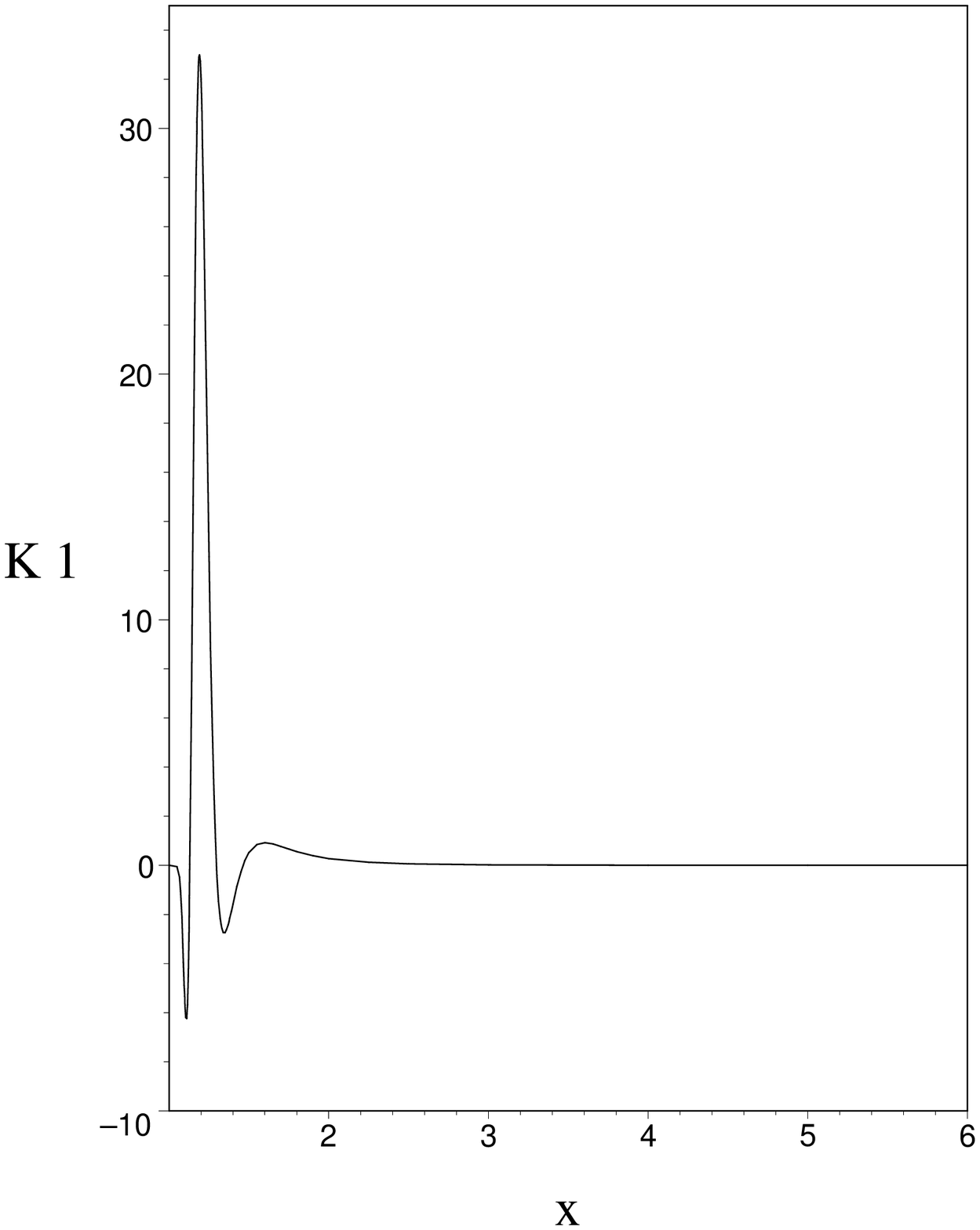}\\[.4cm]
\quad\mbox{(a)}\quad &\quad \mbox{(b)}
\end{array}$\\[.6cm]
\includegraphics[scale=0.35]{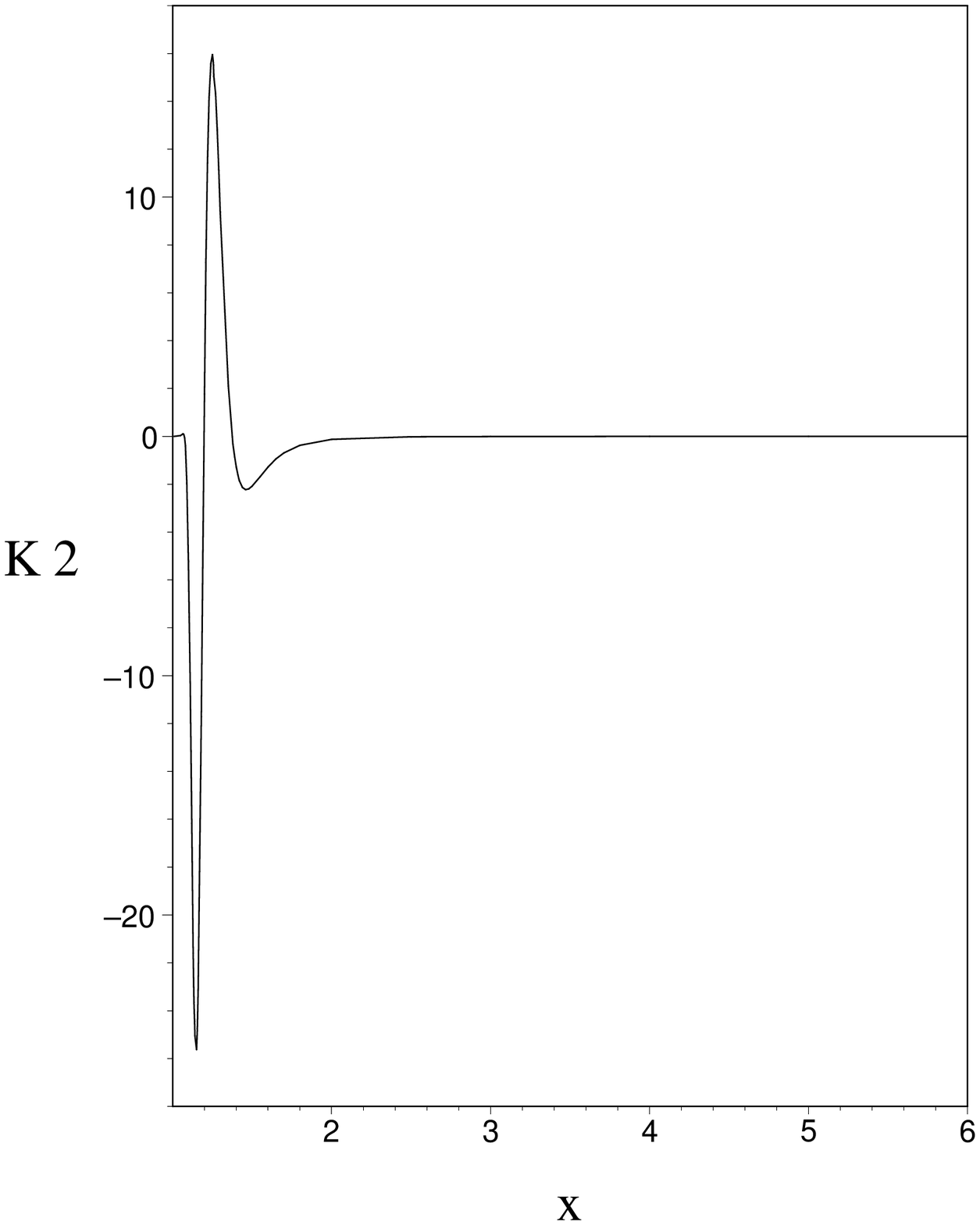}\\[.4cm]
\quad\mbox{(c)}
\end{center}
\caption{The behaviour of the Kretschmann invariant $K_1$ as a function of $x$ is shown for the choice of parameters $a/M=0.5$ and $q=10$ for different values of $y$: (a) $y=0$ and (b) $y=0.5$.
Figure (c) shows instead the behaviour of the Chern-Pontryagin invariant $K_2$ for the same choice of parameters as in Fig. (b). 
Figs. (b) and (c) show how both the invariants $K_1$ and $K_2$ change their signs as $x$ approaches unity. This is a manifestation of the appearence of repulsive gravity regions, typical of naked singularity solutions.
When $x\to\infty$ the invariants $K_1$ and $K_2$ both vanish, as expected.
}
\label{fig:invar}
\end{figure}

In order to investigate the structure of singularities in the QM metric we have to consider the curvature invariants.
Since the solution is a vacuum one there exist just two independent quadratic scalar invariants, the Kretschmann invariant $K_1$ and the Chern-Pontryagin invariant $K_2$ defined by
\beq
K_1=R^{\alpha\beta\gamma\delta}R_{\alpha\beta\gamma\delta}\ , \qquad 
K_2={}^*R^{\alpha\beta\gamma\delta}R_{\alpha\beta\gamma\delta}\ ,
\eeq
where the star denotes the dual.
The behaviour of curvature invariants as functions of $x$ for selected values of $y$, $a/M$ and $q$ is shown in Fig.~\ref{fig:invar}. 
It turns out that $K_1$ diverges when approaching $x=1$ along the direction $y=0$, but it is finite there moving along a different path.
Furthermore, the invariant $K_2$ vanishes identically for $y=0$. 
Therefore, the metric has a directional singularity at $x=1$ (see e.g. \cite{taylor}).

It is now interesting to investigate what kind of hypersurface is $x=1$.
Consider the normal to a $x=\,$ const hypersurface. The behaviour of its norm $g^{xx}$ when approaching $x=1$ discriminates between its character
being either null or timelike depending on the value of the quadrupole parameter. The limit of $g^{xx}$ as $x\to1$ also depends on $y$. 
For instance, approaching $x=1$ along any direction on the equatorial plane $y=0$ gives
\beq
g^{xx}\sim (x-1)^{1+q/2-q^2/4}\ ,
\eeq
implying that the singular hypersurface $x=1$ is null when $|q-1|<\sqrt{5}$ and timelike otherwise.
On the other hand, moving along the axis $y=1$ gives
\beq
g^{xx}\sim\left\{ 
\begin{array}{ll}
(x-1)^{1+q}\ , &\qquad q>0\\
(x-1)\ , &\qquad -1<q<0\\
(x-1)^{-q}\ , &\qquad q<-1\ ,
\end{array}
\right.
\eeq
implying that the singular hypersurface $x=1$ is always null.
The corresponding careful analysis for the Erez-Rosen solution ($a=0$) is contained in Appendix A of \cite{masquev95}.

A similar discussion can be done also for the metric determinant.
Approaching $x=1$ along any direction on the equatorial plane $y=0$ gives
\beq
\sqrt{-g}\sim (x-1)^{-q/2+q^2/4}\ ,
\eeq
implying that the volume element vanishes approaching the singular hypersurface $x=1$ when $q<0$, $q>2$ and diverges otherwise.
On the other hand, moving along the axis $y=1$ gives
\beq
\sqrt{-g}\sim\left\{ 
\begin{array}{ll}
(x-1)^{-q}\ , &\qquad q>0\\
{\rm const}\ , &\qquad -1<q<0\\
(x-1)^{q+1}\ , &\qquad q<-1\ .
\end{array}
\right.
\eeq
We note that if the presence of the quadrupole moment totally changes the situation with respect to the Kerr spacetime, the smooth horizon of the Kerr solution becoming a singular hypersurface in the QM solution, the properties of the naked singularity are also different with respect to the limiting case of the Erez-Rosen spacetime.

\begin{figure} 
\typeout{*** EPS figure ReS}
\begin{center}
\includegraphics[scale=0.35]{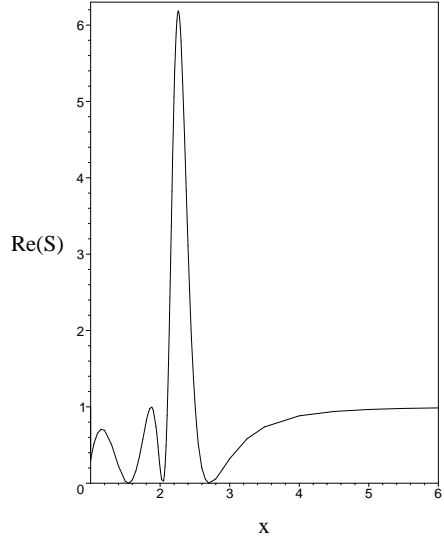}
\end{center}
\caption{The behaviour of the real part of the speciality index $S$ as a function of $x$ is shown for the choice of parameters $a/M=0.5$, $q=10$ and $y=0$. 
In this case ($y=0$) we have also Im$(S)\equiv0$.
}
\label{fig:ReS}
\end{figure}

Finally, the spectral index $S$ (see \ref{NPquant} for the definition of $S$ in terms of Weyl scalars) shows that the solution is algebraically general. 
The real part of $S$ is plotted in Fig.~\ref{fig:ReS} as a function of $x$ for $y=0$ and selected values of $a/M$ and $q$. The imaginary part is identically zero in this case.
A numerical analysis of the spectral index for different values of the quadrupole moment shows that $S\to1$ as $x\rightarrow \infty$, i.e. it is
algebraically special. We conclude that the asymptotic behavior of the spacetime is dominated by the Kerr spacetime which is algebraically special 
of type D. This  is in agreement with the expectation which follows from the analysis of relativistic multipole moments, according to which any stationary axisymmetric asymptotically flat vacuum solution of Einstein's equations must approach the Kerr metric asymptotically \cite{cos80}. 

Note that drawing a Penrose diagram would help to easier understand the global aspects of the geometry of the QM solution. However, due to the rather involved form of the metric functions it is a very hard task to construct it.
Analytic calculations can be performed only in the case of small values of the quadrupole parameter $q$.
But in this simplest case ($q\ll 1$) it is possible to show that the resulting conformal diagram is closely similar to the corresponding one for a Kerr naked singularity (see e.g. Fig.~6.4 of Ref. \cite{carter}).


\section{Geodesics}
\label{sec:geod}

The geodesic motion of test particles is governed by the following equations:
\begin{eqnarray}\fl
\label{geoeqns}
\dot t&=&\frac{E}{f}+\frac{\omega f}{\sigma^2X^2Y^2}(L-\omega E)\ , \qquad 
\dot \phi=\frac{f}{\sigma^2X^2Y^2}(L-\omega E)\ , \nonumber\\
\fl
\ddot y&=&-\frac12\frac{Y^2}{X^2}\left[\frac{f_y}{f}-2\gamma_y+\frac{2y}{X^2+Y^2}\right]{\dot x}^2
+\left[\frac{f_x}{f}-2\gamma_x-\frac{2x}{X^2+Y^2}\right]{\dot x}{\dot y}\nonumber\\
\fl
&&+\frac12\left[\frac{f_y}{f}-2\gamma_y-\frac{2y}{X^2+Y^2}\frac{X^2}{Y^2}\right]{\dot y}^2
-\frac12\frac{e^{-2\gamma}}{f\sigma^4X^2Y^2(X^2+Y^2)}
\left\{Y^2[f^2(L-\omega E)^2\right.\nonumber\\
\fl
&&\left.+E^2\sigma^2X^2Y^2]f_y+2(L-\omega E)f^3[y(L-\omega E)-EY^2\omega_y]\right\}\ , \nonumber\\
\fl
{\dot x}^2&=&-\frac{X^2}{Y^2}{\dot y}^2+\frac{e^{-2\gamma}X^2}{\sigma^2(X^2+Y^2)}\left[E^2-\mu^2f-\frac{f^2}{\sigma^2X^2Y^2}(L-\omega E)^2\right]\ , 
\end{eqnarray}
where Killing symmetries and the normalization condition have been used.
Here $E$ and $L$ are the energy and angular momentum of the test particle respectively, $\mu$ is the particle mass and dot denotes differentiation with respect to the affine parameter; furthermore, the notation 
\beq
X=\sqrt{x^2-1}\ , \quad Y=\sqrt{1-y^2}\ 
\eeq
has been introduced.

Let us consider the motion on the symmetry plane $y=0$.
If $y=0$ and $\dot y=0$ initially, Eq. (\ref{geoeqns})$_3$ ensures that the motion will be confined on the symmetry plane, since $f_y$, $\omega_y$ and $\gamma_y$ all vanish at $y=0$, so that $\ddot y=0$ too.
Eqs. (\ref{geoeqns}) thus reduce to
\begin{eqnarray}\fl\quad
\label{geoeqnsequat}
\dot t&=&\frac{E}{f}+\frac{\omega f}{\sigma^2X^2}(L-\omega E)\ , \qquad 
\dot \phi=\frac{f}{\sigma^2X^2}(L-\omega E)\ , \nonumber\\
\fl\quad
{\dot x}^2&=&\frac{e^{-2\gamma}X^2}{\sigma^2(1+X^2)}\left[E^2-\mu^2f-\frac{f^2}{\sigma^2X^2}(L-\omega E)^2\right]\ , 
\end{eqnarray}
where metric functions are meant to be evaluated at $y=0$.
The motion turns out to be governed by the effective potential $V$ defined by the equation 
\beq
\label{eqV}
V^2-\mu^2f-\frac{f^2}{\sigma^2X^2}(L-\omega V)^2=0\ .
\eeq
In fact, for $E=V$ the rhs of Eq. (\ref{geoeqnsequat})$_3$ vanishes.

\begin{figure} 
\typeout{*** EPS figure invar}
\begin{center}
$\begin{array}{cc}
\includegraphics[scale=0.5]{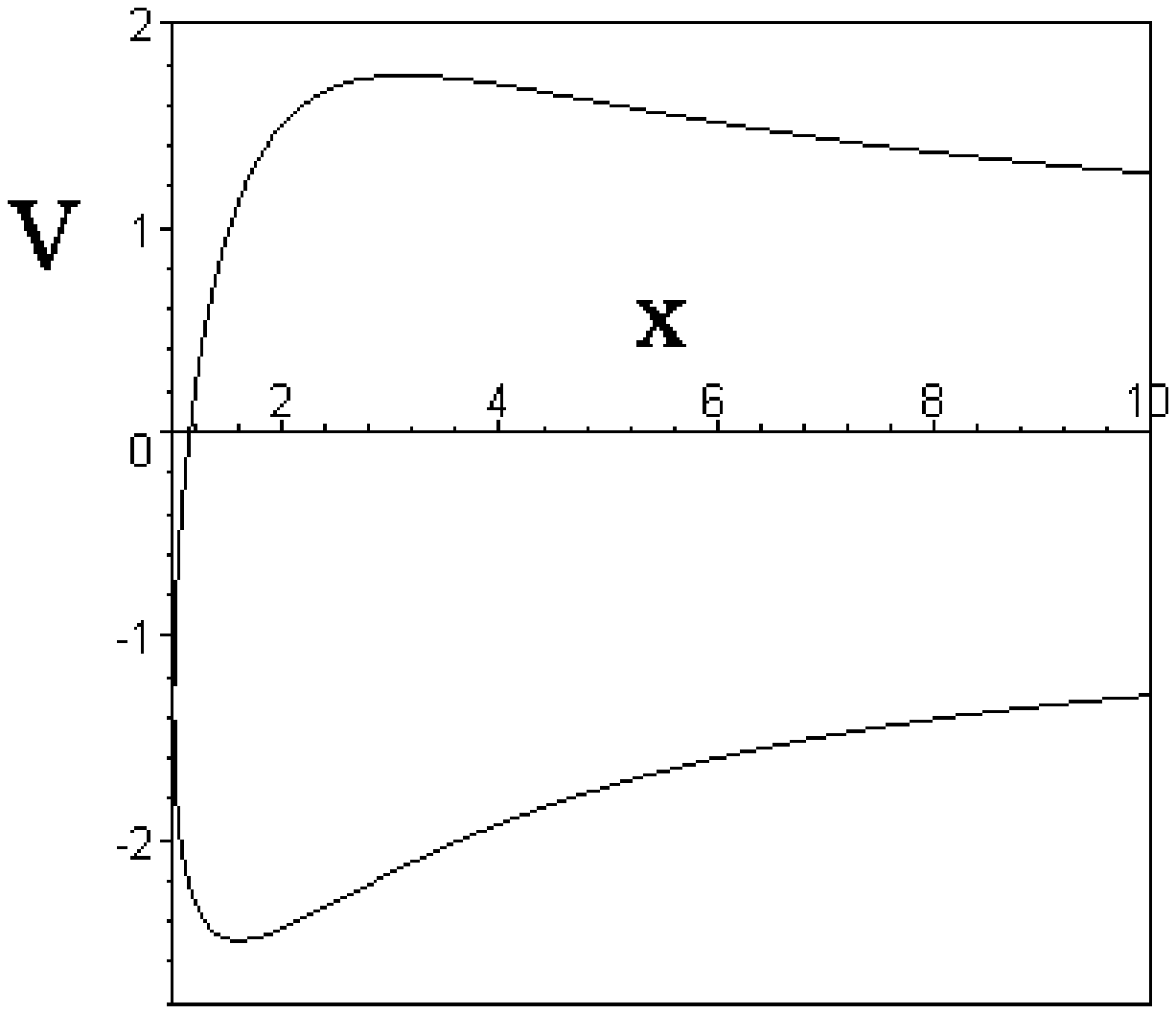}&\quad
\includegraphics[scale=0.5]{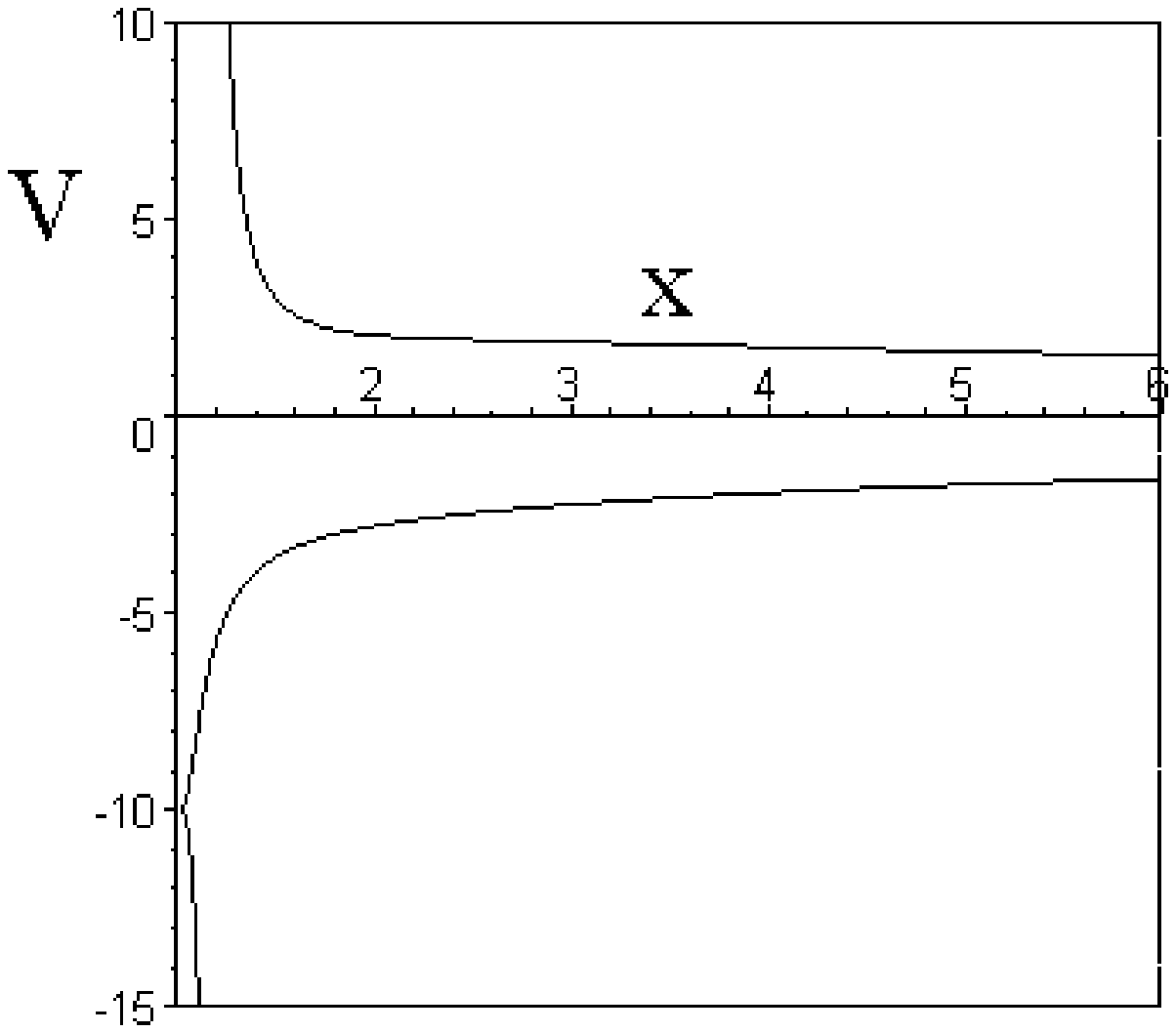}\\[.4cm]
\quad\mbox{(a)}\quad &\quad \mbox{(b)}
\end{array}$\\[.6cm]
\includegraphics[scale=0.45]{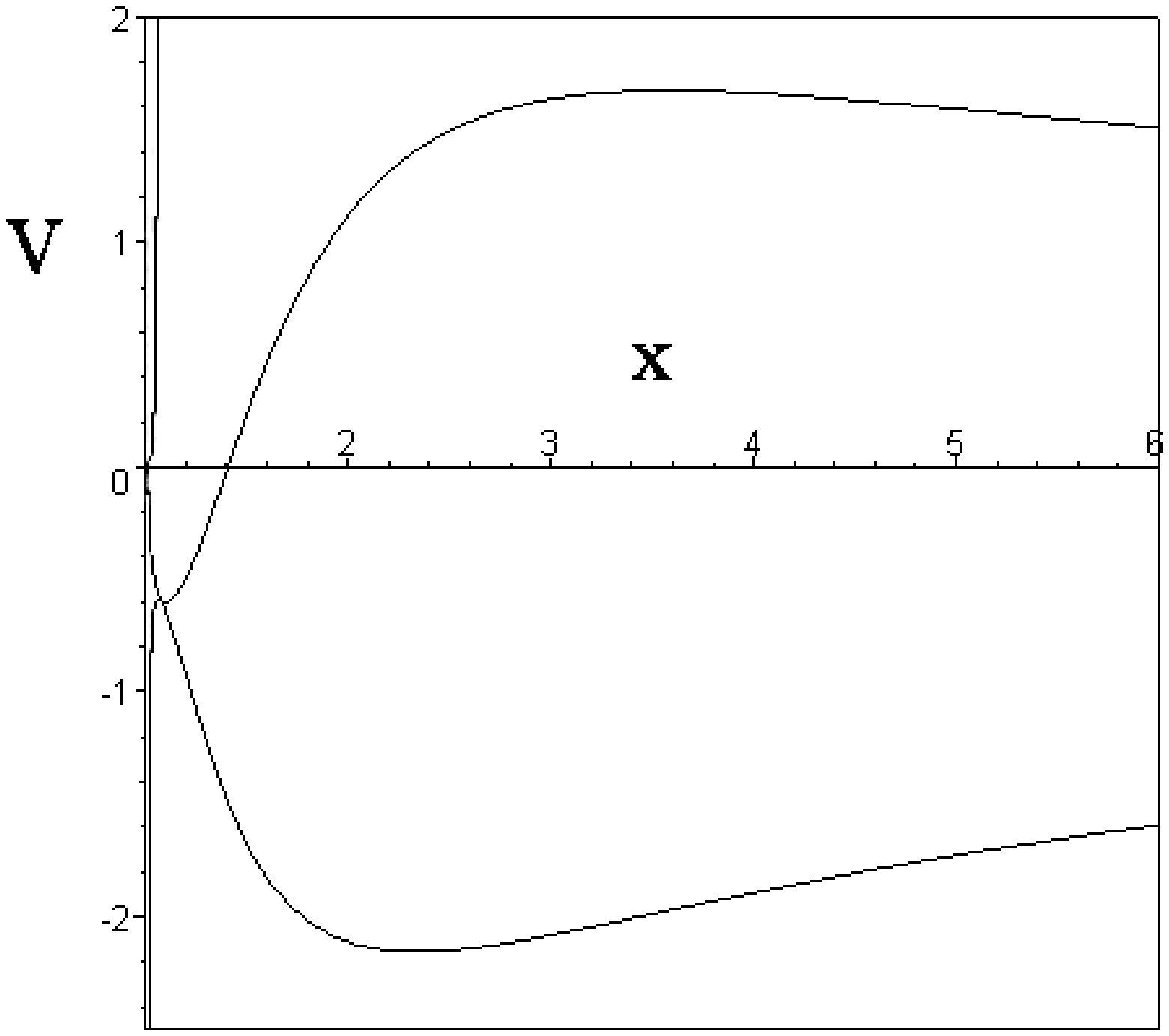}\\[.4cm]
\quad\mbox{(c)}
\end{center}
\caption{The behaviour of the effective potential $V$ as a function of $x$ is shown for the choice of parameters $a/M=0.5$ and $L/(\mu M)=10$ for different values of the quadrupole parameter: (a) $q=0$, (b) $q=10$ and (c) $q=-10$.
}
\label{fig:Veff}
\end{figure}

The behaviour of $V$ as a function of $x$ is shown in Fig.~\ref{fig:Veff}.
Repulsive effects occur for decreasing values of $x$ approaching $x=1$.  
It is worth to mention that the repulsive effect of the naked singularity has also been investigated by Herrera \cite{herrera} in the case of a quasispherical spacetime but without taking into account the rotation of the source.

The case of a geodesic particle at rest will be analyzed below.
Circular geodesics will be discussed in detail in the next section, where accelerated orbits are also studied.


\subsection{Particle at rest}

For the QM solution which is characterized by the presence of a naked singularity it is even possible to satisfy the conditions for a geodesic particle to be at rest for certain values of the quadrupole moment.
This effect can be explained when the attractive behaviour of gravity is balanced by a repulsive force exerted by the naked singularity.
Usually, repulsive effects are interpreted as a consequence of 
the presence of an effective mass which varies with distance and can thus become negative. 
The consideration of the corresponding post-Newtonian limit shows that an effective mass can be indeed
introduced, depending on the distance from the source and the value of the Geroch-Hansen quadrupole 
moment \cite{quev90}. We see that in the case of the corresponding exact solution under consideration a similar situation takes place.

\begin{figure} 
\typeout{*** EPS figure equil}
\begin{center}
$\begin{array}{cc}
\includegraphics[scale=0.4]{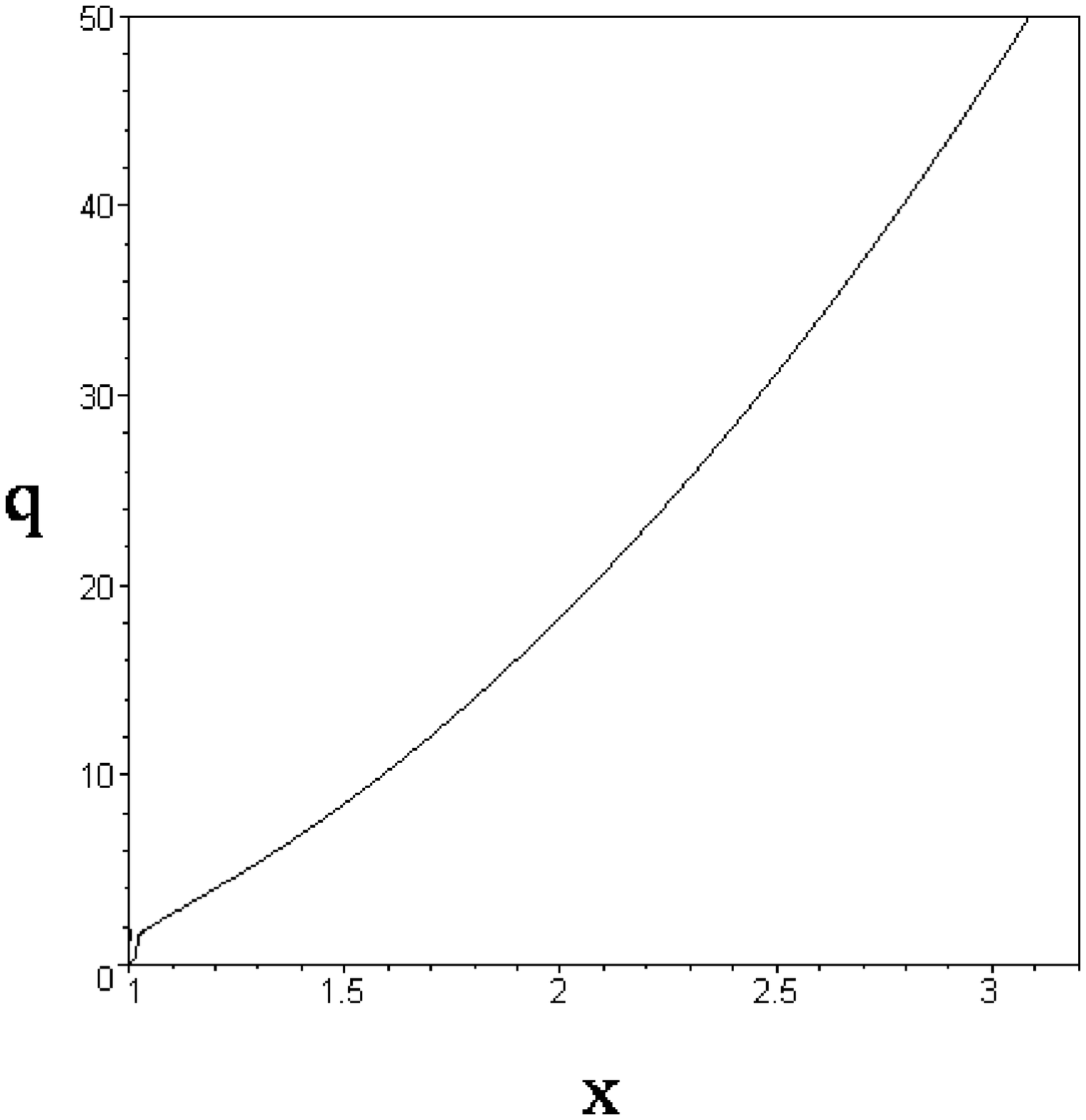}&\quad
\includegraphics[scale=0.4]{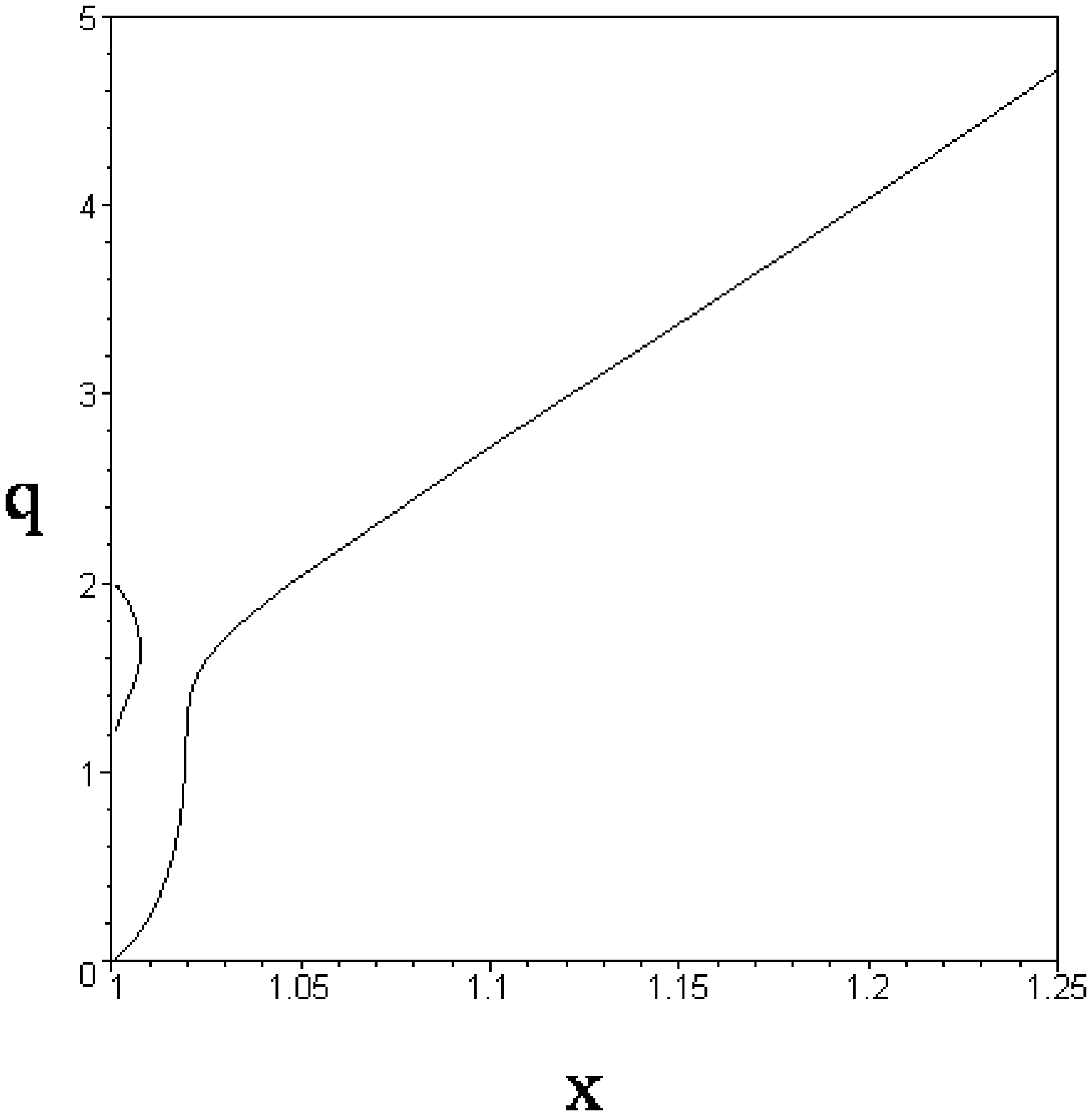}\\[.4cm]
\quad\mbox{(a)}\quad &\quad \mbox{(b)}
\end{array}$\\
\end{center}
\caption{The pairs $(x,q)$ allowing a test particle to be at rest in the QM spacetime are shown for $a/M=0.5$.
Fig. (b) is a detail of Fig. (a) close to $x=1$.
No equilibrium positions exist for $q\leq0$.
}
\label{fig:equil}
\end{figure}

A particle at rest is characterized by the four velocity
\beq
U=\frac{1}{\sqrt{f}}\partial_t\ .
\eeq
The corresponding four acceleration $a(U)=\nabla_UU$ is given by
\beq
a(U)=\frac{e^{-2\gamma}}{2\sigma^2(X^2+Y^2)}[X^2f_x\partial_x+Y^2f_y\partial_y]\ .
\eeq
On the symmetry plane $y=0$ we have $f_y=0$, so that the geodesic condition $a(U)=0$ implies $f_x=0$.
The pairs $(x,q)$ satisfying this condition are shown in Fig.~\ref{fig:equil} for a fixed value of $a/M$.
As an example, for $a/M=0.5$ and $q=10$ we get $x\approx1.588$ as the equilibrium position.
In this case the corresponding energy and angular momentum per unit mass of the particle are given by $E/\mu\approx  0.553$ and $L/(\mu M)\approx0.257$ respectively.
No equilibrium positions exist for $q\leq0$.


\section{Circular orbits on the symmetry plane}
\label{sec:cir}

Let us introduce the ZAMO family of fiducial observers, with four velocity
\beq
\label{n}
n=N^{-1}(\partial_t-N^{\phi}\partial_\phi);
\eeq
here $N=(-g^{tt})^{-1/2}$ and $N^{\phi}=g_{t\phi}/g_{\phi\phi}$ are the lapse and shift functions respectively. A suitable orthonormal frame adapted to  ZAMOs is given by
\beq
e_{\hat t}=n , \,\quad
e_{\hat x}=\frac1{\sqrt{g_{xx}}}\partial_x, \,\quad
e_{\hat y}=\frac1{\sqrt{g_{yy}}}\partial_y, \,\quad
e_{\hat \phi}=\frac1{\sqrt{g_{\phi \phi }}}\partial_\phi ,
\eeq
with dual
\beq\fl\quad
\omega^{{\hat t}}=N\rmd t\ , \quad \omega^{{\hat x}}=\sqrt{g_{xx}}\rmd x\ , \quad
\omega^{{\hat y}}= \sqrt{g_{yy}} \rmd y\ , \quad
\omega^{{\hat \phi}}=\sqrt{g_{\phi \phi }}(\rmd \phi+N^{\phi}\rmd t)\ .
\eeq

The 4-velocity $U$ of uniformly rotating circular orbits
can be parametrized either by the (constant) angular velocity with respect to infinity $\zeta$ or, equivalently, by the (constant) linear velocity  $\nu$ with respect to ZAMOs
\beq
\label{orbita}
U=\Gamma [\partial_t +\zeta \partial_\phi ]=\gamma [e_{\hat t} +\nu e_{\hat \phi}], \qquad \gamma=(1-\nu^2)^{-1/2}\ ,
\eeq
where $\Gamma$ is a normalization factor which assures that $U_\alpha U^\alpha =-1$  given by:
\beq
\Gamma =\left[ N^2-g_{\phi\phi}(\zeta+N^{\phi})^2 \right]^{-1/2}=\frac{\gamma}{N}
\eeq
 and
\beq
\zeta=-N^{\phi}+\frac{N}{\sqrt{g_{\phi\phi}}} \nu .
\eeq
We limit our analysis to the motion on the symmetry plane ($y=0$) of the solution (\ref{metgen})--(\ref{metquev}).
Note both $y=0$ and $x=x_0$ are constants along any given circular orbit, and that the azimuthal coordinate along the orbit depends on the coordinate time $t$ or proper time $\tau$ along that orbit according to
\beq\label{eq:phitau}
  \phi -\phi_0 = \zeta t = \Omega_U \tau_U \ ,\quad
\Omega_U =\Gamma\zeta
\ ,
\eeq
defining the corresponding coordinate and proper time orbital angular velocities $\zeta$ and $\Omega_U$. These determine the rotation of the spherical frame with respect to a nonrotating frame at infinity.

The spacetime Frenet-Serret frame along a single timelike test particle worldline with 4-velocity $U=E_0$ and parametrized by the proper time $\tau_U$ is described by the following system of evolution equations \cite{iyer-vish}
\begin{eqnarray}
\label{FSeqs}
\frac{DE_0}{d\tau_U}&=\kappa E_1\ , \qquad &
\frac{DE_1}{d\tau_U}=\kappa E_0+\tau_1 E_2\ ,\nonumber \\
 \nonumber \\
\frac{DE_2}{d\tau_U}&=-\tau_1E_1+\tau_2E_3\ , \qquad &
\frac{DE_3}{d\tau_U}=-\tau_2E_2\ .
\end{eqnarray}
The absolute value of the curvature $\kappa$ is the magnitude of
the acceleration $a(U)\equiv DU/d\tau_U=\kappa E_1$, while
the first
and second torsions $\tau_1$ and $\tau_2$ are the components of the Frenet-Serret angular velocity vector
\beq
\label{omegaFS}
\omega_{\rm (FS)}=\tau_1 E_3 + \tau_2 E_1\ , \qquad ||\omega_{\rm (FS)}||=[\tau_1^2 + \tau_2^2]^{1/2}\ ,
\eeq
with which the spatial Frenet-Serret frame $\{E_a\}$ rotates with respect to a Fermi-Walker transported frame along $U$.
It is well known that any circular orbit on the symmetry plane of a reflection symmetric spacetime has zero second torsion $\tau_2$, while the geodesic curvature $\kappa$ and the first torsion $\tau_1$ are simply related by
\beq
\tau_1= -\frac{1}{2\gamma^2} \frac{\rmd \kappa}{\rmd \nu}\ .
\eeq

On the symmetry plane there exists a large variety of special
circular orbits  \cite{bjdf,idcf1,idcf2,bjm}; particular interest
is devoted to the co-rotating $(+)$ and counter-rotating $(-)$
timelike circular geodesics whose linear velocities is
\beq\fl\quad
\label{nugeo}
\nu_{({\rm geo})\, \pm}\equiv \nu_\pm =
\frac{fC\pm\left[f^2\omega^2-\sigma^2(x^2-1)\right]\sqrt{D}}{\sqrt{x^2-1}\sigma
\{f_x[f^2\omega^2+\sigma^2(x^2-1)]+2f(f^2\omega\omega_x-\sigma^2x)\}}
\ ,
\eeq
where
\begin{eqnarray}
C&=&-2\sigma^2(x^2-1)\omega f_x-f\{\omega_x[f^2\omega^2+\sigma^2(x^2-1)]-2\sigma^2x\omega\}\ , \nonumber\\
D&=&f^4\omega_x^2-\sigma^2f_x[f_x(x^2-1)-2xf]\ .
\end{eqnarray}
All quantities in the previous expressions are meant to be evaluated at $y=0$.
The corresponding timelike conditions $|\nu_\pm|<1$ together with the reality condition $D\geq0$ identify the
allowed regions for the \lq\lq radial'' coordinate where co/counter-rotating
geodesics exist. 

Other special orbits correspond to the \lq\lq geodesic
meeting point observers'' defined in \cite{idcf2}, with
\beq
\nu_{\rm (gmp)}= \frac{\nu_{+}+\nu_{-}}{2}\ .
\eeq

A Frenet-Serret (FS) intrinsic frame along $U$ \cite{iyer-vish} is given by
\begin{eqnarray}\fl\quad
E_0\equiv U=\gamma [n+\nu e_{\hat \phi}]\ , \quad E_1=e_{\hat x}\ , \quad E_2=e_{\hat y}\ , \quad E_2\equiv E_{\hat \phi}=\gamma [\nu n+e_{\hat \phi}]\ .
\end{eqnarray}
It is also convenient to introduce the Lie relative curvature of each orbit \cite{idcf2}
\begin{eqnarray}\fl\quad
k_{\rm (lie)}&=&-\partial_{\hat x} \ln \sqrt{g_{\phi\phi}}\nonumber\\
\fl\quad
&=&\frac{e^{-\gamma}\sqrt{x^2-1}}{2\sigma x\sqrt{f}}\left\{\frac{\sigma^2[(x^2-1)f_x-2xf]+f^2\omega(2f\omega_x+\omega f_x)}{\sigma^2(x^2-1)-f^2\omega^2}\right\}\ .
\end{eqnarray}
It then results
\begin{eqnarray}
\label{ketau1}
\kappa &=&k_{\rm (lie)}\gamma^2 (\nu-\nu_+)(\nu-\nu_-)\ , \nonumber \\
\tau_1&=& k_{\rm (lie)}\nu_{\rm (gmp)}\gamma^2 (\nu-\nu_{{\rm (crit)}+})(\nu-\nu_{{\rm (crit)}-})\ ,
\end{eqnarray}
where
\beq
\nu_{{\rm (crit)}\pm}=\frac{\gamma_- \nu_- \mp \gamma_+ \nu_+}{\gamma_- \mp \gamma_+}\ , \qquad 
\nu_{{\rm (crit)}+}\nu_{{\rm (crit)}-}=1\ ,
\eeq
identify the so called \lq\lq extremely accelerated observers'' \cite{idcf2,fdfacc}: $\nu_{{\rm (ext)}}\equiv\nu_{{\rm (crit)}-}$, which satisfies the timelike condition in the regions where timelike geodesics exist, while $\nu_{{\rm (crit)}+}$ is always spacelike there.

\begin{figure} 
\typeout{*** EPS figure 1ab}
\begin{center}
$\begin{array}{cc}
\includegraphics[scale=0.4]{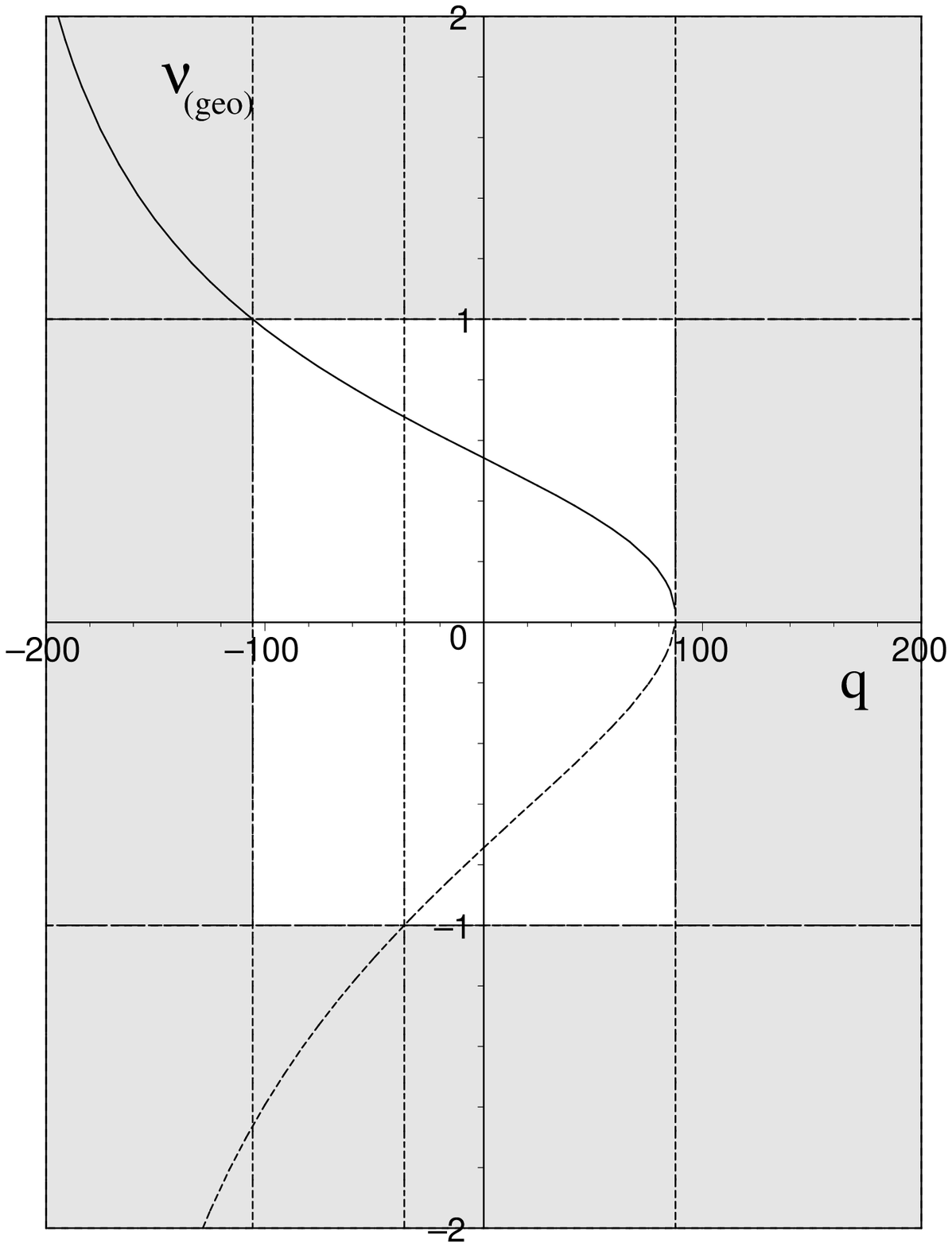}&\qquad
\includegraphics[scale=0.4]{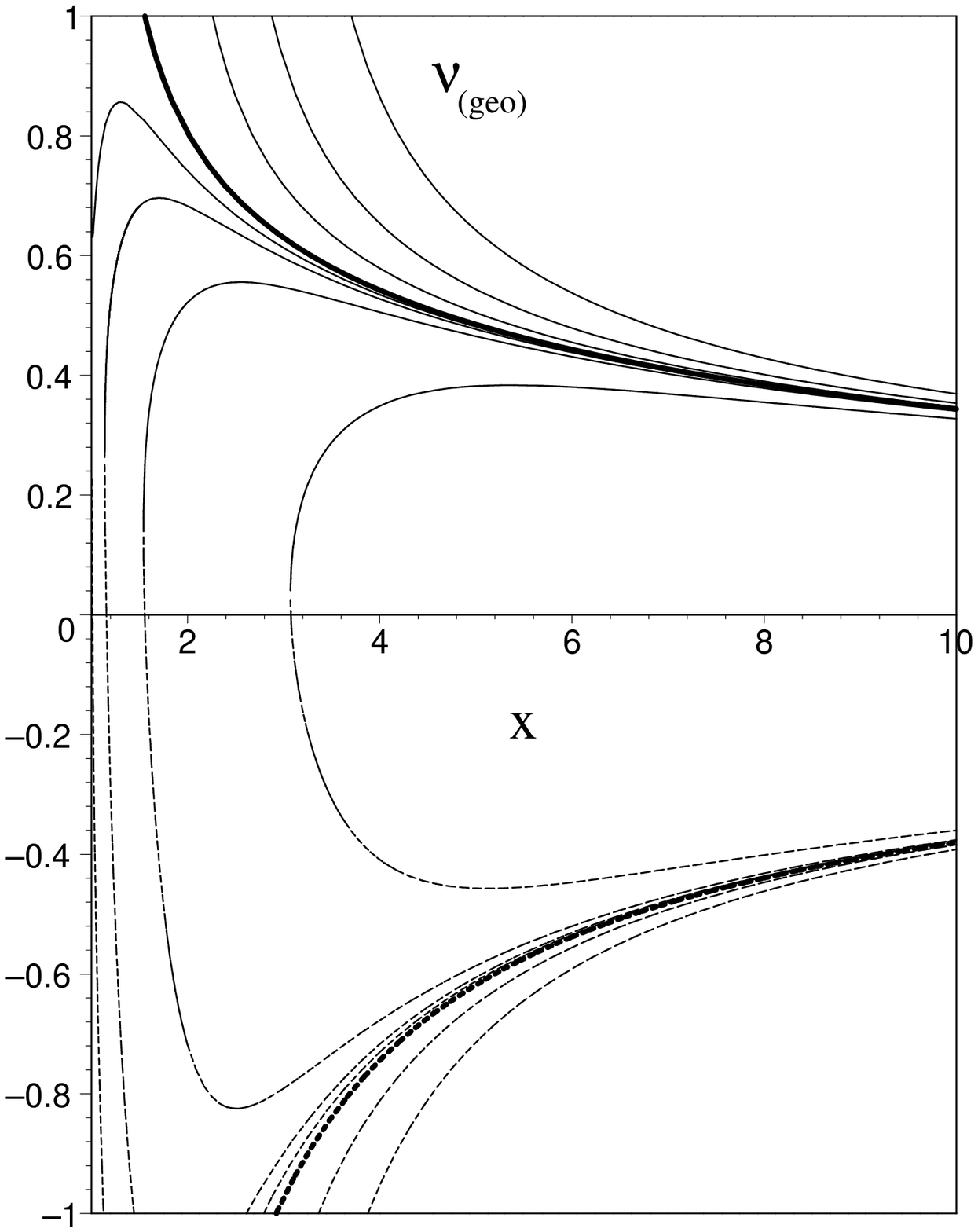}\\[0.4cm]
\mbox{(a)} & \mbox{(b)}\\
\end{array}$
\end{center}
\caption{The geodesic linear velocities $\nu_{\rm (geo)}^{\pm}$ are plotted in Fig. (a) as functions of the quadrupole parameter $q$ for fixed distance parameter $x=4$ from the source and $a/M=0.5$. 
Co-rotating and counter-rotating circular geodesics exist for $q_1<q<q_3$ and $q_2<q<q_3$ respectively, with $q_1\approx-105.59$, $q_2\approx-36.29$ and $q_3\approx87.68$ for this choice of parameters.
The behaviour of $\nu_{\rm (geo)}^{\pm}$ as functions of $x$ is shown in Fig. (b) for different values of $q=[-80,-30,-10,0,2,4,10,50]$.
The thick curves correspond to the Kerr case ($q=0$). 
Curves corresponding to great positive value of the quadrupole parameter in the allowed range exhibit both a local maximum ($\nu_{\rm (geo)}^{+}$) and a local mimimum ($\nu_{\rm (geo)}^{-}$). 
For decreasing values of $q$ the local minimum first disappears; for $q$ further decreasing also the local maximum disappears (the curves are thus ordered from left to right for increasing values of $q$).
Curves corresponding to negative values of $q$ never present extrema as in the case of Kerr spacetime ($q=0$), the lightlike condition being reached at greater values of the \lq\lq radial'' distance for decreasing values of $q$.
}
\label{fig:1}
\end{figure}

The geodesic velocities (\ref{nugeo}) are plotted in Fig.~\ref{fig:1} both as functions of the quadrupole parameter $q$ for fixed \lq\lq radial'' distance (see Fig. (a)) and as functions of $x$ for different values of $q$ (see Fig. (b)).
In the first case (Fig. (a)) we have shown how the quadrupole moment affects the causality condition: there exist a finite range of values of $q$ wherein timelike circular geodesics are allowed: $q_1<q<q_3$ for co-rotating and $q_2<q<q_3$ for counter-rotating circular geodesics.
The critical values $q_1$, $q_2$ and $q_3$ of the quadrupole parameter can be (numerically) determined from Eq. (\ref{nugeo}).  
The difference from the Kerr case is clear instead from Fig. (b): the behaviour of the velocities differs significantly at small distances from the source, whereas it is quite similar for large distances. 

\begin{figure} 
\typeout{*** EPS figure 2ab}
\begin{center}
$\begin{array}{cc}
\includegraphics[scale=0.4]{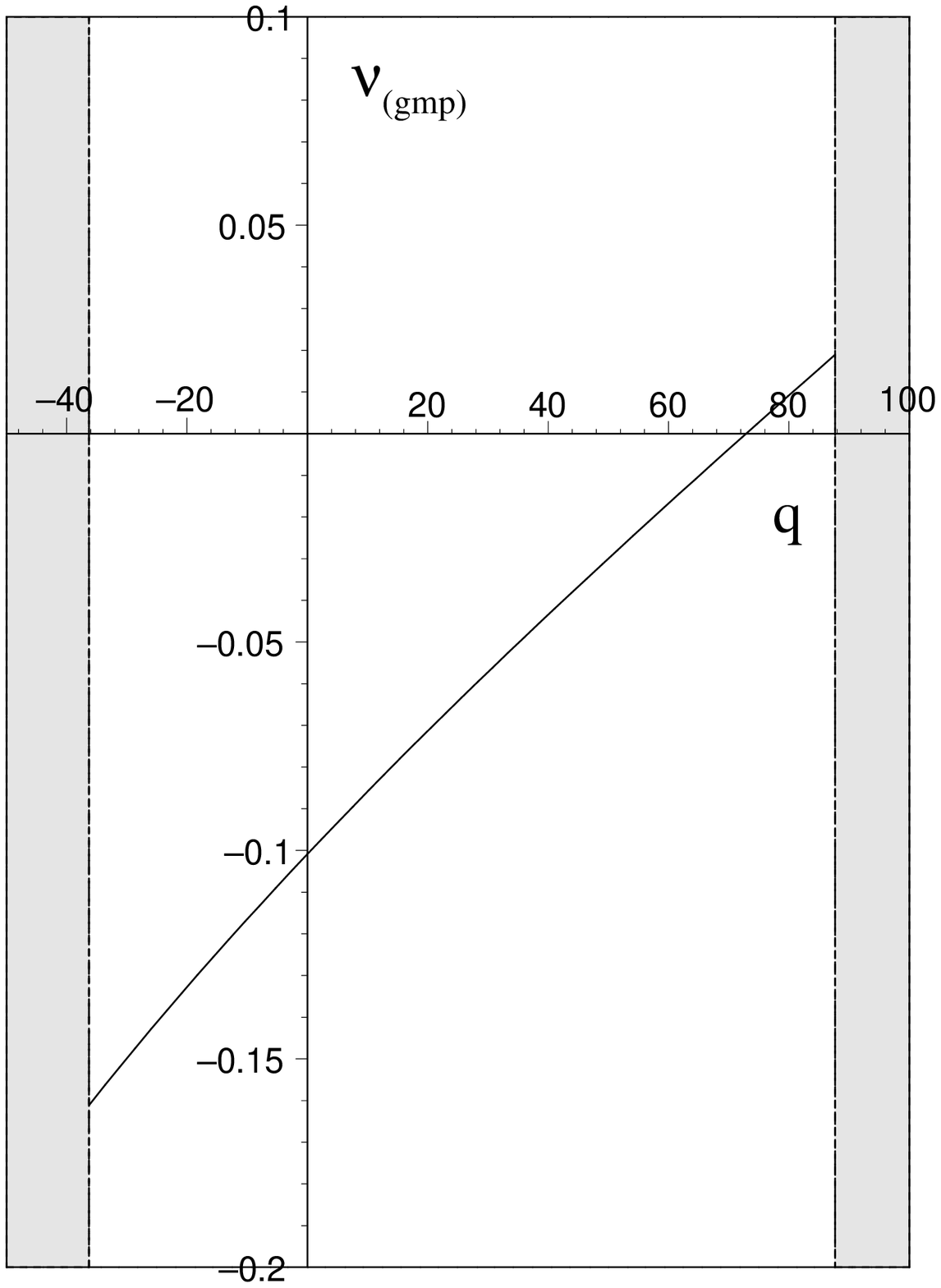}&\qquad
\includegraphics[scale=0.4]{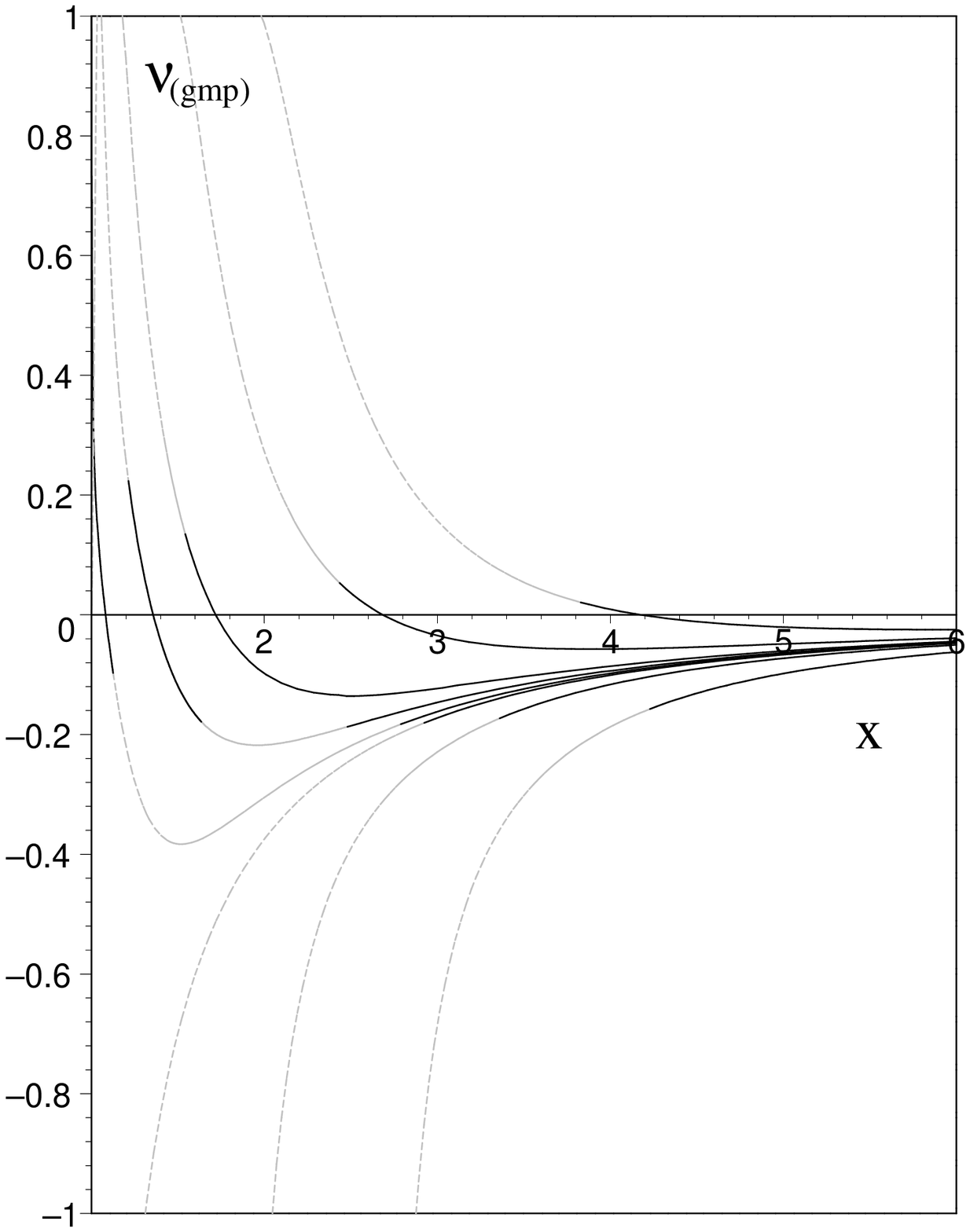}\\[0.4cm]
\mbox{(a)} & \mbox{(b)}\\
\end{array}$
\end{center}
\caption{The linear velocity $\nu_{\rm (gmp)}$ corresponding to the \lq\lq geodesic meeting point observers'' is plotted in Fig. (a) as a function of the quadrupole parameter $q$ for fixed distance parameter $x=4$ from the source and $a/M=0.5$. 
The behaviour of $\nu_{\rm (gmp)}$ as a function of $x$ is shown in Fig. (b) for different values of $q=[-50,-10,0,2,5,10,30,80]$.
It exists only in those ranges of $x$ where both $\nu_{\rm (geo)}^{+}$ and $\nu_{\rm (geo)}^{-}$ exist (solid black lines). 
These ranges are listed below:
$q=-50$, $x\gtrsim4.23$;
$q=-10$, $x\gtrsim3.36$;
$q=0$, $x\gtrsim2.92$;
$q=2$, $1.0012\lesssim x\lesssim1.13$ and $x\gtrsim2.79$;
$q=5$, $1.21\lesssim x\lesssim1.64$ and $x\gtrsim2.48$;
$q=10$, $x\gtrsim1.54$;
$q=30$, $x\gtrsim2.44$;
$q=80$, $x\gtrsim3.83$.
}
\label{fig:2}
\end{figure}

\begin{figure} 
\typeout{*** EPS figure 3ab}
\begin{center}
$\begin{array}{cc}
\includegraphics[scale=0.4]{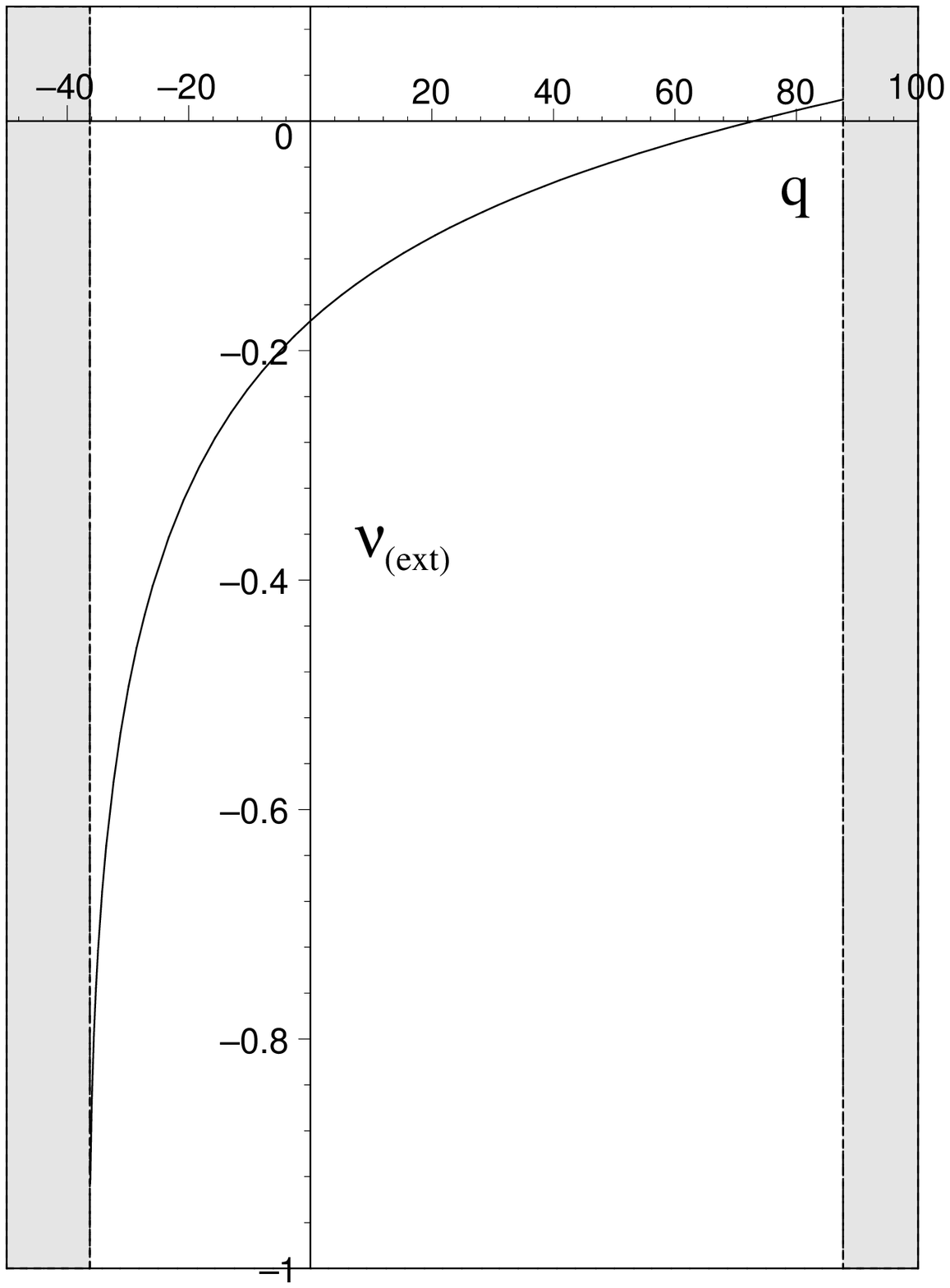}&\qquad
\includegraphics[scale=0.4]{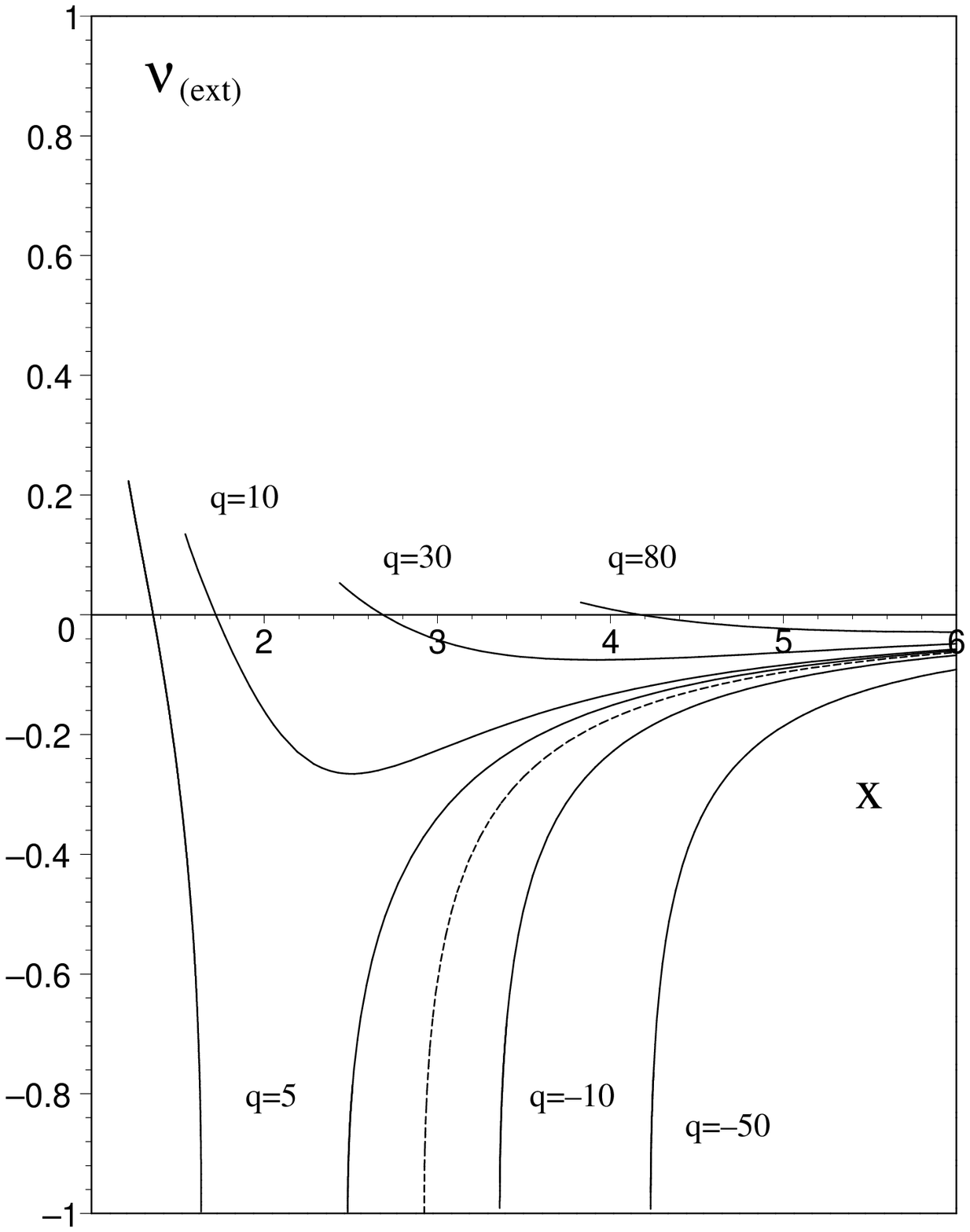}\\[0.4cm]
\mbox{(a)} & \mbox{(b)}\\
\end{array}$
\end{center}
\caption{The linear velocity $\nu_{{\rm (ext)}}$ corresponding to the \lq\lq extremely accelerated observers'' is plotted in Fig. (a) as a function of the quadrupole parameter $q$ for fixed distance parameter $x=4$ from the source and $a/M=0.5$. 
The behaviour of $\nu_{{\rm (ext)}}$ as a function of $x$ is shown in Fig. (b) for different values of $q=[-50,-10,0,5,10,30,80]$. The dashed curve corresponds to the case $q=0$.
It exists only in those ranges of $x$ where both $\nu_{\rm (geo)}^{+}$ and $\nu_{\rm (geo)}^{-}$ exist (see Fig. \ref{fig:2}). 
}
\label{fig:3}
\end{figure}

A similar discussion concerning the linear velocity $\nu_{\rm (gmp)}$ of \lq\lq geodesic meeting point observers'' as well as $\nu_{{\rm (ext)}}$ of \lq\lq extremely accelerated observers'' is done in Figs.~\ref{fig:2} and~\ref{fig:3} respectively.


\begin{figure} 
\typeout{*** EPS figure kab}
\begin{center}
$\begin{array}{cc}
\includegraphics[scale=0.4]{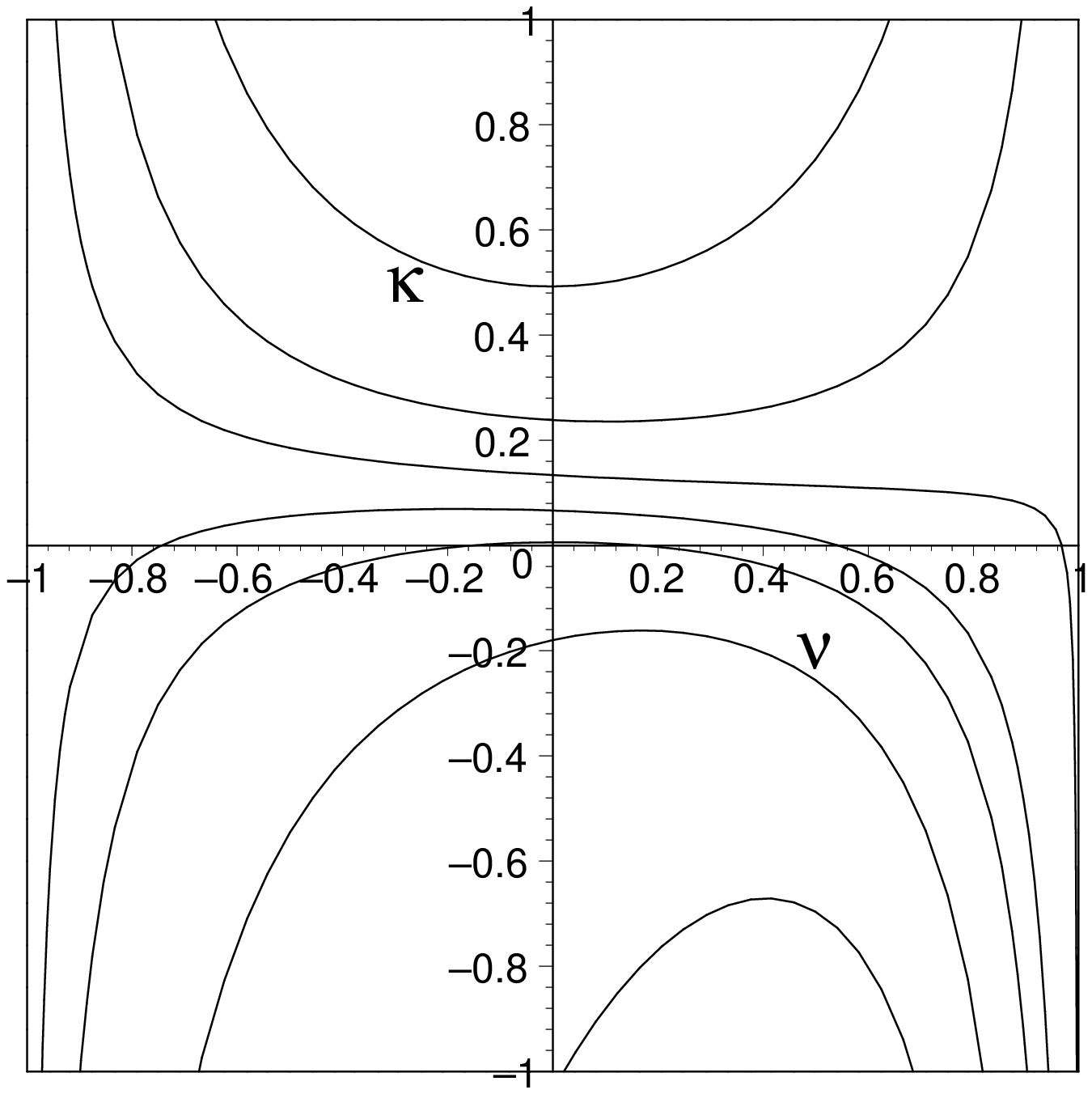}&\quad
\includegraphics[scale=0.4]{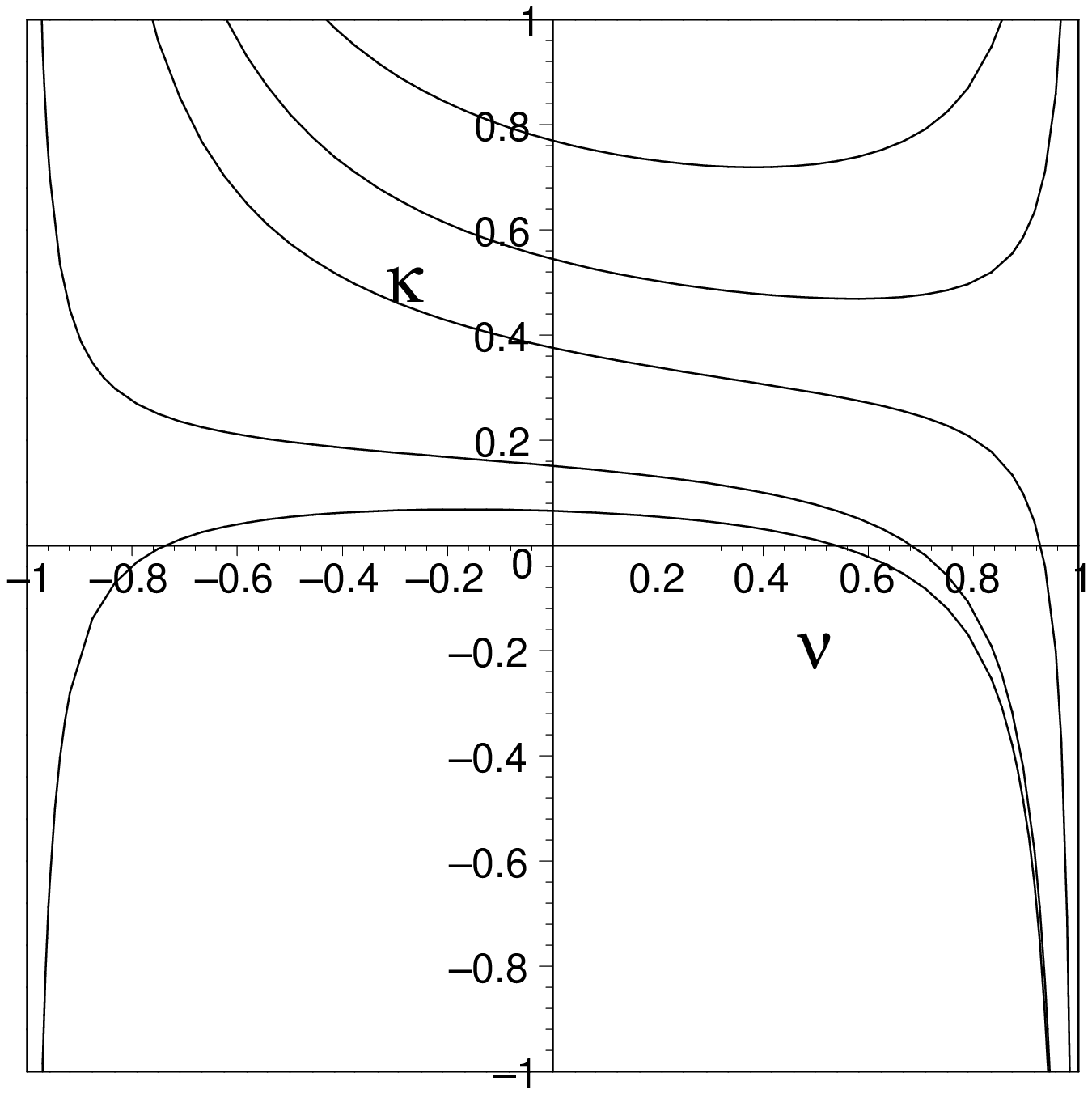}\\[.4cm]
\quad\mbox{(a)}\quad &\quad \mbox{(b)}
\end{array}$\\[.6cm]
\includegraphics[scale=0.4]{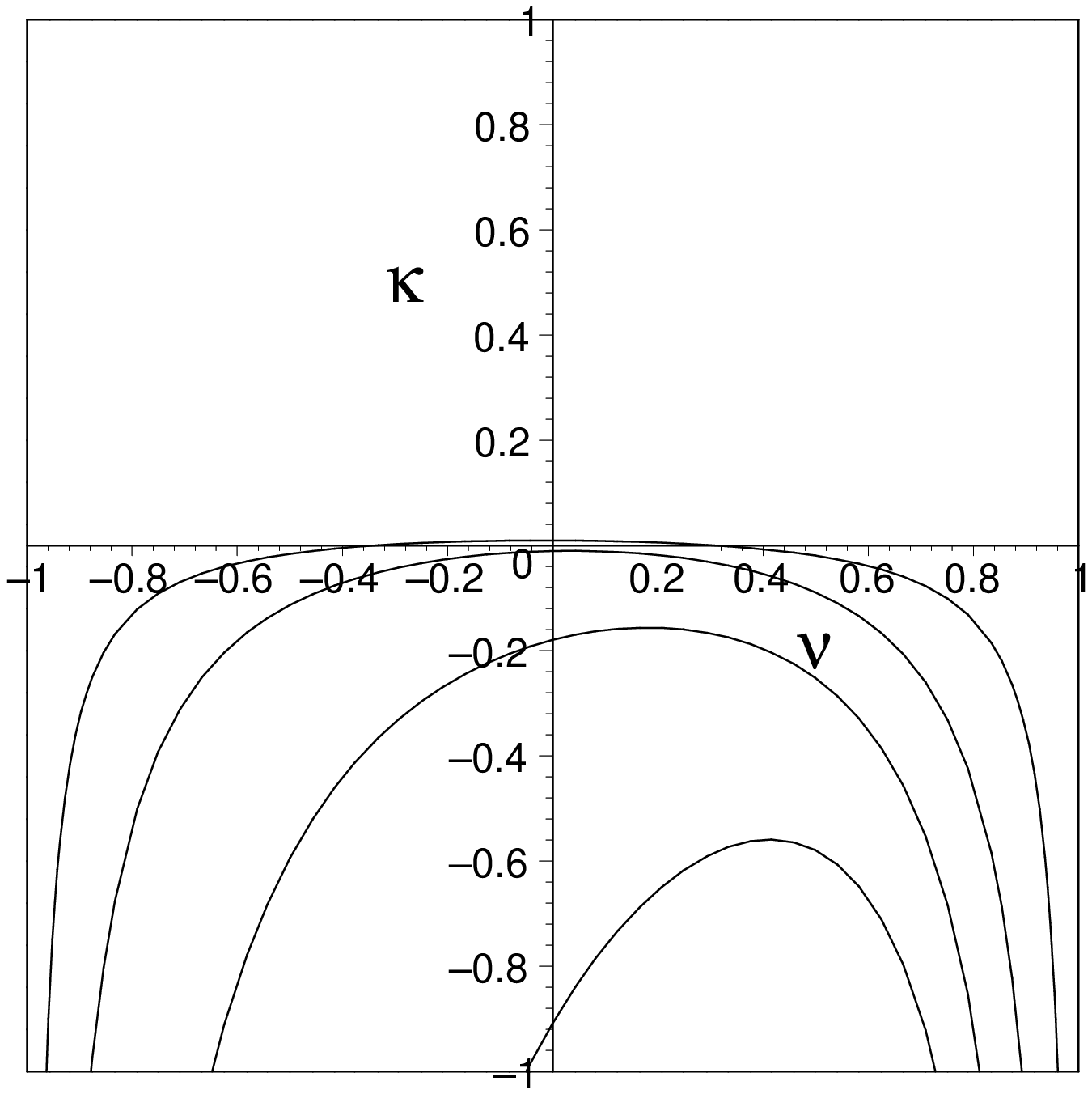}\\[.4cm]
\quad\mbox{(c)}
\end{center}
\caption{The acceleration $\kappa$ for circular orbits at $y=0$ is plotted in Fig. (a) as a function of $\nu$ for $a/M=0.5$, $x=4$ and different values of the quadrupole parameter: $q=[-500,-250,-100,0,80,250,500]$. 
The curves are ordered from top to bottom for increasing values of $q$. 
The values of $\nu$ associated with $\kappa=0$ correspond to geodesics, i.e. $\nu_{\rm (geo)}^{\pm}$. 
The behaviour of $\kappa$ as a function of $\nu$ for different $x$ is shown in Figs. (b) and (c) for fixed values of the quadrupole parameter: (b) $q=1$, $x=[1.1,1.25,1.5,2.5,4]$ and (c) $q=100$, $x=[2.5,3,4,10]$.  
}
\label{fig:k}
\end{figure}

\begin{figure} 
\typeout{*** EPS figure t1ab}
\begin{center}
$\begin{array}{cc}
\includegraphics[scale=0.4]{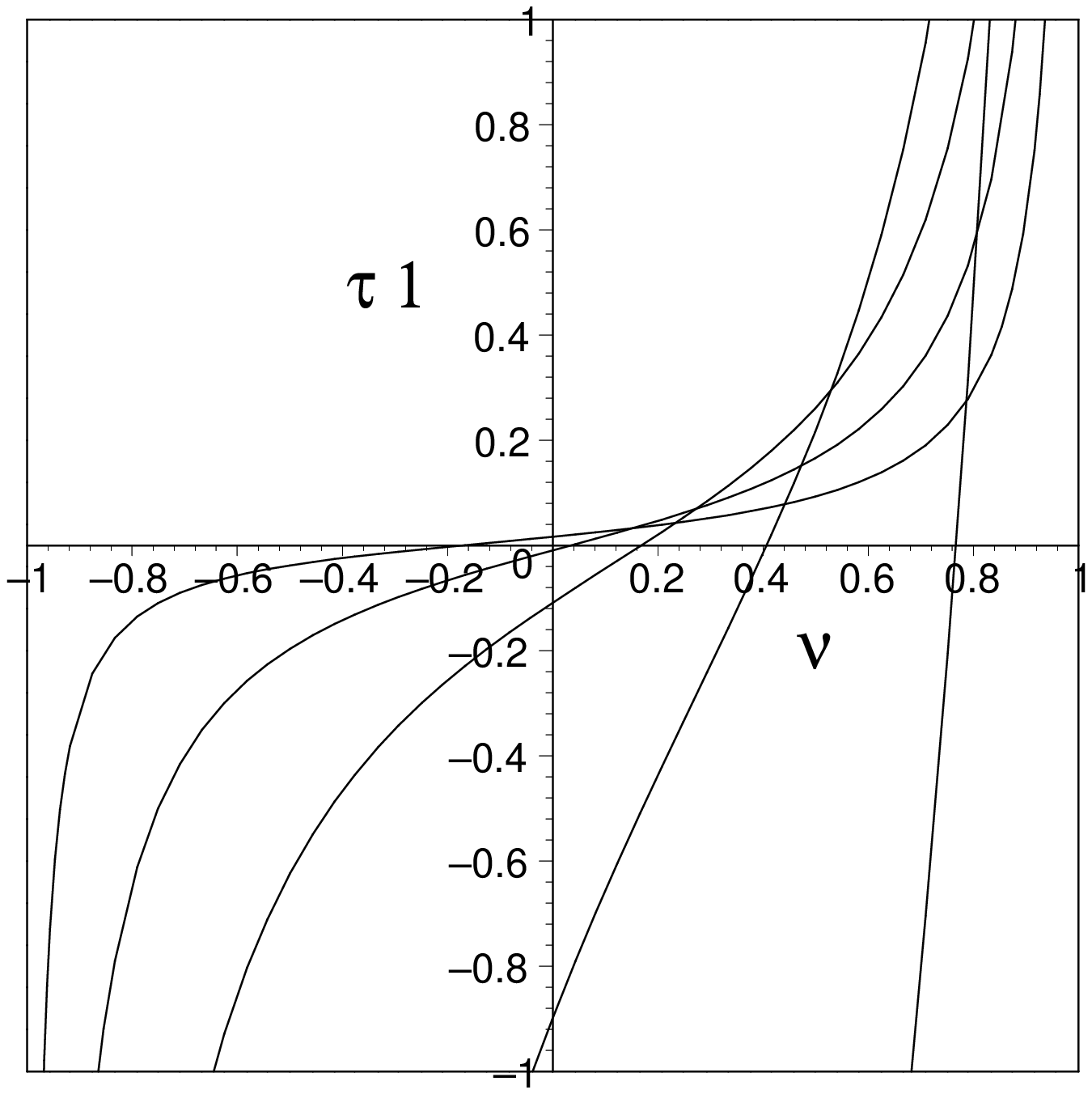}&\quad
\includegraphics[scale=0.4]{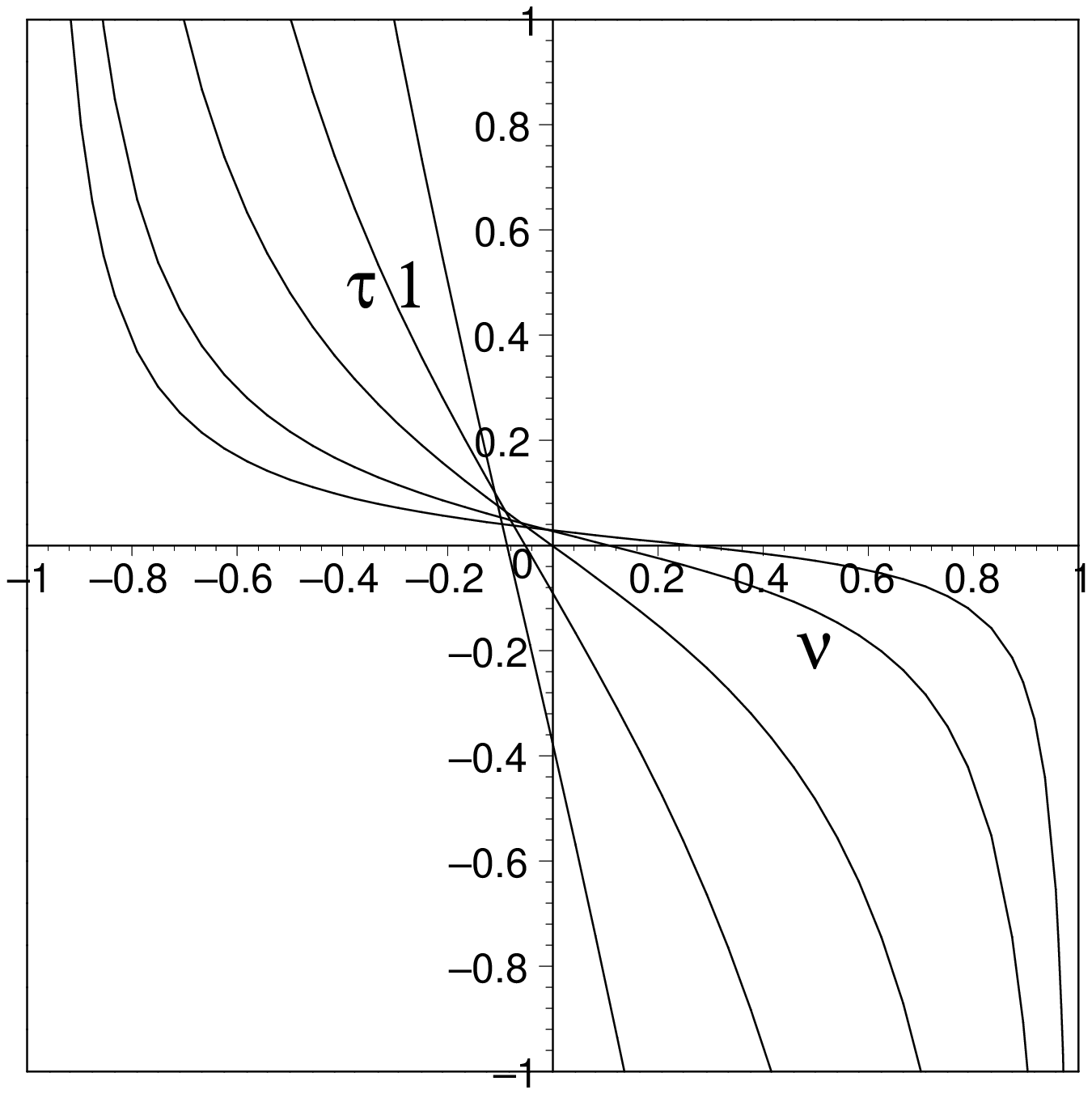}\\[.4cm]
\quad\mbox{(a)}\quad &\quad \mbox{(b)}
\end{array}$\\[.6cm]
\includegraphics[scale=0.4]{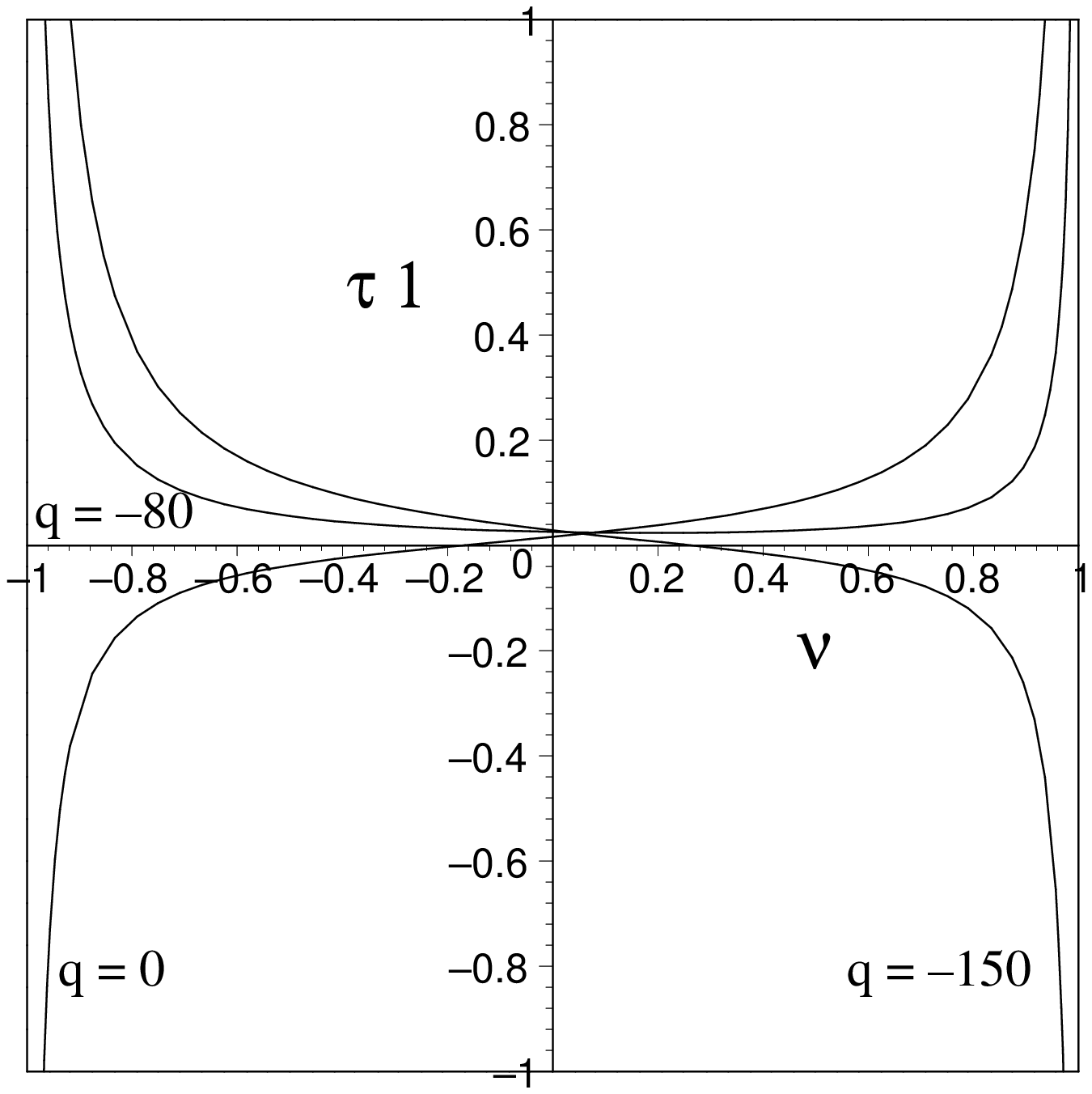}\\[.4cm]
\quad\mbox{(c)}
\end{center}
\caption{The first torsion $\tau_1$ for circular orbits at $y=0$ is plotted as a function of $\nu$ for $a/M=0.5$, $x=4$ and different values of the quadrupole parameter: (a) $q=[0,100,250,500,750]$, (b) $q=[-1000,-750,-500,-250,-150]$ and (c) $q=[-150,-80,0]$.
The curves are ordered from left to right for increasing/decreasing values of $q$ in Fig. (a)/Fig. (b) respectively.
Figure (c) shows the changes of behavior occurring in the different regions $q<q_1$, $q_1<q<q_2$ and $q>q_2$ (see Fig. \ref{fig:1}). 
}
\label{fig:t1}
\end{figure}

The behaviours of both the acceleration $\kappa$ and the first torsion $\tau_1$ as functions of $\nu$ are shown in Figs.~\ref{fig:k} and~\ref{fig:t1} for different values of the quadrupole parameter and fixed $x$ as well as for different values of the \lq\lq radial'' distance and fixed $q$. 
A number of interesting effects do occur.
For instance, from Fig.~\ref{fig:k} (a) and (b) one recognizes certain counter-intuitive behaviours of the acceleration, in comparison with our Newtonian experience. These effects have their roots also in the Kerr solution and are well known and studied since the 90s \cite{idcf2}. For instance, for negative values of $q$ ($q=-500, -250$) we see that by increasing the speed $\nu$ (for positive values over that corresponding to the local minimum) the acceleration also increases; hence, in order to maintain the orbit, the particle (a rocket, say) should accelerate outwards. This is counter-intuitive in the sense that for a circular orbit at a fixed radius increasing the speed corresponds to an increase of the centrifugal acceleration and therefore to maintain the orbit \lq\lq classically" one would expect to supply an acceleration inward. All such effects have been 
analyzed in the past decades in the Kerr spacetime in function of the radius of the orbit, i.e. the distance from the black hole.
The novelty here is represented by Fig.~\ref{fig:k} (a) where the various curves do not correspond to different orbital radii (i.e. different values of the coordinate $x$, as it is for the cases (b) and (c)), but to different values of the quadrupole parameter $q$ at a fixed radial distance.
Therefore, the conclusion is that a spacecraft orbiting around an extended body --according to general relativity and the QM solution-- should expect counter-intuitive engine acceleration to remain on a given orbit. Outward or inward extra acceleration for an increase of the speed critically depend on the quadrupole moment (i.e. the physical structure) of the source, a fact that should be taken into account.

Fig.~\ref{fig:t1} shows instead that at a fixed radius  one can always find a value of the quadrupole parameter and a value of the speed at which the first torsion vanishes. Being the second torsion identically vanishing, in these conditions the Frenet-Serret frame becomes also a Fermi-Walker frame: in fact the Frenet-Serret angular velocity represent the rate of rotation of the Frenet-Serret frame with respect to a Fermi-Walker one.
In \cite{idcf2} (see Sec. 6, Eq. (6.22)) the precession of a test gyroscope (Fermi-Walker dragged along a circular orbit) has been related to the first torsion of an observer-adapted Frenet-Serret frame. The same discussion  repeated here together with a simple inspection of  Fig.~\ref{fig:t1} allows us to include the effects of the presence of the quadrupole parameter $q$.


\section{Concluding remarks}

In this work we investigated some properties of the QM spacetime which is a generalization of Kerr spacetime, 
including an arbitrary mass quadrupole moment. Our results show that  a deviation from spherical symmetry, corresponding to a non-zero gravitoelectric quadrupole moment, completely changes the structure of spacetime. A similar behaviour has been found in the case of the Erez-Rosen spacetime \cite{masquev95}. 
A naked singularity appears that affects the ergosphere and introduces regions where closed timelike curves are allowed. Whereas in the Kerr spacetime the ergosphere corresponds to the boundary of a simply-connected region of spacetime, in the present case the ergosphere is distorted by the presence of the quadrupole and can even become transformed into multiply-connected regions. All these changes occur near the naked singularity.

The presence of a naked singularity leads to interesting consequences in the motion of test particles. For instance, repulsive effects can take place
in a region very close to the naked singularity. In that region stable circular orbits can exist.  The limiting case of 
static particle is also allowed, due to the balance of the gravitational attraction and the repulsive force exterted by the naked singularity. 

We have studied the family of circular orbits on the symmetry plane of the QM solution, analyzing all their relevant intrinsic properties, namely Frenet-Serret curvature and torsions. We have also selected certain special circular orbits, like the \lq\lq geodesic meeting points'' orbits (i.e. orbits which contain the meeting points of two oppositely rotating circular geodesics) and the \lq\lq extremely accelerated'' orbits (i.e. orbits with respect to which the relative velocities of two oppositely rotating circular geodesics are opposite), whose kinematical characterization was given in the 90s \cite{idcf2} with special attention to Kerr spacetime. Here we have enriched their  properties specifying the dependence on the quadrupole parameter.

The question about the stability of the QM solution is important for astrophysical purposes. In this context, we have obtained some preliminary results by using the variational formulation of the perturbation problem as developed explicitly by Chandrasekhar \cite{chandra} for stationary  axisymmetric solutions.  
A numerical analysis performed for fixed values of the parameters entering the QM metric shows that it is unstable against perturbations that preserve axial symmetry.  
One can indeed expect that, once an instability sets in, the final state of gravitational collapse will be described by the Kerr spacetime, the multipole moments of the initial configuration decaying during the black hole formation.  
Nevertheless, a more detailed analysis is needed in order to completely establish the stability properties of this solution.

Finally, we mention the fact that it is possible to generalize the metric investigated in this work to include the case of a non spherically symmetric mass distribution endowed with 
an electromagnetic field \cite{quevmas90}. The resulting exact solution of Einstein-Maxwell equations turns out to be asymptotically flat, contains the Kerr-Newman black hole spacetime as a special case, and is characterized by two infinite sets of gravitational and electromagnetic multipole moments.  
For a particular choice of the parameters, the solution is characterized by the presence of  a naked singularity. It would be interesting to explore 
repulsive effects generated by the electromagnetic field of the naked singularity also in this case. This task will be treated in a future work.

\section*{Acknowledgements}

The authors are indebted to Prof. B. Mashhoon and Prof. R. Ruffini for useful discussions. 
All thank IcraNet for support.

\appendix

\section{General form of QM solution with arbitrary Zipoy-Voorhees parameter}
\label{QMgeneral}

The general form of QM solution with arbitrary Zipoy-Voorhees parameter $\delta$ is given by Eq. (\ref{metgen}) with functions 
\begin{eqnarray}
f&=&\frac{R}{L} e^{-2q\delta P_2Q_2}\ , \nonumber\\
\omega&=&-2a-2\sigma\frac{\mathfrak M}{R} e^{2q\delta P_2Q_2}\ , \nonumber\\
e^{2\gamma}&=&\frac{1}{4}\left(1+\frac{M}{\sigma}\right)^2\frac{R}{(x^2-1)^\delta} e^{2\delta^2\hat\gamma}\ ,
\end{eqnarray}
where $\hat\gamma$ is the same as in Eq. (\ref{variousdefs}), while 
\begin{eqnarray}
R&=& a_+ a_- + b_+b_-\ , \qquad 
L = a_+^2 + b_+^2\ , \nonumber\\
{\mathfrak M}&=&(x+1)^{\delta-1}[x(1-y^2)(\lambda+\mu) a_+ +y(x^2-1)(1-\lambda\mu)b_+]\ .
\end{eqnarray}
The functions $a_\pm$ and $b_\pm$ are now given by
\begin{eqnarray}
a_\pm &=& (x\pm1)^{\delta-1}[x(1-\lambda\mu) \pm (1+\lambda\mu)]\ , \nonumber\\
b_\pm &=& (x\pm1)^{\delta-1}[x(\lambda+\mu) \mp (\lambda-\mu)]\ ,
\end{eqnarray}
with 
\begin{eqnarray}
\lambda &=& \alpha (x^2-1)^{1-\delta}(x+y)^{2\delta-2}e^{2q\delta\delta_+}\ , \nonumber\\
\mu &=& \alpha (x^2-1)^{1-\delta}(x-y)^{2\delta-2}e^{2q\delta\delta_-}\ .
\end{eqnarray}
The functions $\delta_\pm$ and the constants $\alpha$ and $\sigma$ are instead the same as in Eqs. (\ref{variousdefs2}) and (\ref{metquev}) respectively.

This solution reduces to the solution (\ref{metgen})--(\ref{metquev}) for $\delta=1$.

\section{Newman-Penrose quantities}
\label{NPquant}

Let us adopt here the metric signature $(+,-,-,-)$ in order to use the Newman-Penrose formalism in its original form and then easily get the necessary physical quantities \cite{chandra}.
The Weyl-Lewis-Papapetrou metric is thus given by
\begin{eqnarray}\fl\quad
\label{metgen2}
\rmd s^2&=&f(\rmd t-\omega \rmd\phi)^2\nonumber\\
\fl\quad
&&-\frac{\sigma^2}{f}\left\{e^{2\gamma}\left(x^2-y^2\right)\left(\frac{\rmd x^2}{x^2-1}+\frac{\rmd y^2}{1-y^2}\right)+(x^2-1)(1-y^2)\rmd\phi^2\right\}\ ,
\end{eqnarray}

Introduce the following tetrad 
\begin{eqnarray}
\label{NPframe}
l&=&\sqrt{\frac{f}{2}}\left[\rmd t-\left(\omega+\frac{\sigma XY}{f}\right)\rmd\phi\right]\ , \nonumber\\
n&=&\sqrt{\frac{f}{2}}\left[\rmd t-\left(\omega-\frac{\sigma XY}{f}\right)\rmd\phi\right]\ , \nonumber\\
m&=&\frac{1}{\sqrt{2}}\frac{\sigma}{\sqrt{f}}e^{\gamma}\sqrt{X^2+Y^2}\left[\frac{\rmd x}{X}+i\frac{\rmd y}{Y}\right]\ ,
\end{eqnarray}
where 
\beq
X=\sqrt{x^2-1}\ , \quad Y=\sqrt{1-y^2}\ .
\eeq
The nonvanishing spin coefficients are
\begin{eqnarray}\fl\quad
\kappa&=&-{\mathcal A}\left[-\frac{XY}{f}(Xf_x+iYf_y)+\frac{f}{\sigma}(X\omega_x+iY\omega_y)+xY-iyX\right]\ , \nonumber\\
\fl\quad
\tau&=&-\pi^{*}={\mathcal A}(xY-iyX)\ , \nonumber\\
\fl\quad
\nu&=&-{\mathcal A}\left[\frac{XY}{f}(Xf_x-iYf_y)+\frac{f}{\sigma}(X\omega_x-iY\omega_y)-xY-iyX\right]\ , \nonumber\\
\fl\quad
\alpha&=&{\mathcal A}XY\sigma^4\left[-\frac{1}{2f}(Xf_x-iYf_y)+X\gamma_x-iY\gamma_y-\frac{f}{2\sigma}\left(\frac{\omega_x}{Y}-i\frac{\omega_y}{X}\right)\right.\nonumber\\
\fl\quad
&&\left.+\frac{xX+iyY}{X^2+Y^2}\right]\ , \nonumber\\
\fl\quad
\beta&=&{\mathcal A}XY\sigma^4\left[\frac{1}{2f}(Xf_x+iYf_y)-X\gamma_x-iY\gamma_y-\frac{f}{2\sigma}\left(\frac{\omega_x}{Y}+i\frac{\omega_y}{X}\right)\right.\nonumber\\
\fl\quad
&&\left.-\frac{xX-iyY}{X^2+Y^2}\right]\ , 
\end{eqnarray}
where 
\beq
{\mathcal A}=\frac14\sqrt{2}\frac{e^{-\gamma}}{\sigma XY}\sqrt{\frac{f}{X^2+Y^2}}\ .
\eeq
The nonvanishing Weyl scalars are
\begin{eqnarray}\fl\quad
\psi_0&=&{\mathcal A}XY\left[\tau(Xf_x+iYf_y)+2f(X\kappa_x+iY\kappa_y)-\frac{f^2}{\sigma XY}(\kappa+\tau)(X\omega_x+iY\omega_y)\right]\nonumber\\
\fl\quad
&&+2\kappa\beta+\tau(\kappa-\tau)\ , \nonumber\\
\fl\quad
\psi_2&=&\frac1{6}\frac{{\mathcal A}^2X^2Y^2}{f}\left[x\left(f_x+2f\gamma_x-3i\frac{f^2}{\sigma X^2}\omega_y\right)-y\left(f_y+2f\gamma_y+3i\frac{f^2}{\sigma Y^2}\omega_x\right)\right.\nonumber\\
\fl\quad
&&\left.+3i\frac{f}{\sigma}(f_x\omega_y-f_y\omega_x)-X^2(f_{xx}-2f\gamma_{xx})-Y^2(f_{yy}-2f\gamma_{yy})\right.\nonumber\\
\fl\quad
&&\left.+\frac{f^3}{\sigma^2}\left(\frac{\omega_x^2}{Y^2}+\frac{\omega_y^2}{X^2}\right)\right]\ , \nonumber\\
\fl\quad
\psi_4&=&{\mathcal A}XY\left[-\pi(Xf_x-iYf_y)-2f(X\nu_x-iY\nu_y)-\frac{f^2}{\sigma XY}(\nu+\pi)(X\omega_x-iY\omega_y)\right]\nonumber\\
\fl\quad
&&+2\nu\alpha+\pi(\nu-\pi)\ .
\end{eqnarray}

Finally the two scalar invariants of the Weyl tensor whose ratio defines the
speciality index have the following expressions in terms of the Newman-Penrose
curvature quantities
\beq
I=\psi_0\psi_4+3\psi_2^2\ , \qquad 
J=\psi_0\psi_2\psi_4-\psi_2^3\ , \qquad
S=27\frac{J^2}{I^3}\ .
\eeq
$S$ has the value 1 for algebraically special spacetimes.

\end{document}